\documentclass[reqno,11pt]{article}
\usepackage{ifpdf}
\usepackage[usenames,dvipsnames]{xcolor}
\usepackage{ae}
\usepackage[T1]{fontenc}
\usepackage[ansinew]{inputenc}
\usepackage{mathrsfs}
\usepackage{amsmath}
\usepackage{amssymb}
\usepackage{graphicx}
\usepackage{color}
\definecolor{darkblue}{cmyk}{0.9,0.9,0,0}
\definecolor{darkgreen}{rgb}{0,0.55,0}

\usepackage{hyperref}

\usepackage{epsfig}
\usepackage{amsmath}

\newcommand{\comment}[1]{}

\newcommand{\beq}{\begin{equation}}
\newcommand{\eeq}{\end{equation}}
\newcommand{\beqq}{\begin{equation*}}
\newcommand{\eeqq}{\end{equation*}}
\newcommand\beqa{\begin{eqnarray}}
\newcommand\eeqa{\end{eqnarray}}
\newcommand\beqaa{\begin{eqnarray*}}
\newcommand\eeqaa{\end{eqnarray*}}
\newcommand\bea{\begin{array}}
\newcommand\eea{\end{array}}

\def\XXint#1#2#3{{\setbox0=\hbox{$#1{#2#3}{\int}$ }
\vcenter{\hbox{$#2#3$ }}\kern-.5\wd0}}

\def\XXint#1#2#3{{\setbox0=\hbox{$#1{#2#3}{\int}$}
\vcenter{\hbox{$#2#3$}}\kern-.5\wd0}}

\newcommand{\nn}{\nonumber}

\newcommand{\neqa}{\nonumber\end{eqnarray}}
\newcommand{\la}[1]{\label{#1}}

\newcommand{\eq}[1]{(\ref{#1})}

\def\tr{{\rm tr~}}

\newcommand{\hs}{\frac{\sqrt{3}}{2}}
\renewcommand{\d}{\partial}

\newcommand{\<}{{\langle}}
\renewcommand{\>}{{\rangle}}

\newcommand{\re}{\relax{\rm I\kern-.18em R}}

\renewcommand{\sp}{p\hspace{-.40em}/}

\def\su2{{SU(2)}}

\def\[{\left[}
\def\]{\right]}

\def\s{\sigma}

\def\({\left(}
\def\){\right)}
\def\[{\left[}
\def\]{\right]}

\def\<{\langle}
\def\>{\rangle}

\def\bT{{\bf T}}

\def\bC{{\bf C}}

\def\mC{{\mathbb C}}

\def\s*{\ *_{\!\!\!\!\!\!\!\!\!\,_{\,_\text{\scriptsize{sym}}}}}
\def\hs*{\ \hat{*}_{\!\!\!\!\!\!\!\!\!\,_{\,_\text{\scriptsize{sym}}}}}
\def\d{\partial}

\def\i2{\frac{i}{2}}

\def\spi{\relax{\rm \pi\kern-0.5em /}}
\def\sA{\relax{\rm A\kern-0.5em /}}
\def\sp{\relax{\rm p\kern-0.5em /}}
\def\sd{\relax{\rm \d\kern-0.5em /}}
\def\sk{\relax{\rm k\kern-0.5em /}}
\def\sn{\relax{\rm n\kern-0.5em /}}
\def\sl{\relax{\rm l\kern-0.5em /}}
\def\sP{\relax{\rm P\kern-0.7em /}}
\def\sBethe{\relax{\rm \Bethe\kern-0.5em /}}

\def\bC{{\bold C}}

\newcommand{\bra}[1]{\langle #1 |}
\newcommand{\ket}[1]{| #1 \rangle}

\newcommand{\g}{{\rm {good}}}

\newcommand{\svx}{{\mathsf{x}}}

\newcommand{\svy}{{\mathsf{y}}}

\numberwithin{equation}{section}

\usepackage{varioref}
\usepackage{makeidx}
\makeindex

\usepackage[english]{babel}
\usepackage[numbers,sort&compress,comma]{natbib}

\begin{document}

\begin{titlepage}

\setcounter{page}{0}
\renewcommand{\thefootnote}{\fnsymbol{footnote}}


\vspace{1cm}

\begin{center}

\textbf{\Large\mathversion{bold} Separation of variables for higher rank integrable models:\\ review of recent progress}

\vspace{1cm}

{\large Fedor Levkovich-Maslyuk \footnote{{\it Email:\/}
{\ttfamily fedor.levkovich at gmail.com}}
} 

\vspace{1cm}

\it Centre for Mathematical Science, City St George's, University of London, \\ Northampton Square, EC1V 0HB, London, UK

\vspace{1cm}

{\bf Abstract}
\end{center}
\vspace{-.3cm}
\begin{quote}
Separation of variables (SoV) is a powerful method expected to be applicable for a wide range of quantum integrable systems, from models in condensed matter physics to gauge and string theories. Yet its full implementation for many higher rank examples, such as $SU(N)$ spin chains with $N>2$, has remained elusive for a long time. In this pedagogical review we discuss the major progress achieved recently in understanding SoV for models of this type. In particular, for rational $SU(N)$ spin chains we describe different constructions of the SoV basis, novel compact forms for spin chain eigenstates, the functional SoV approach, and explicit computation of the SoV measure. We also discuss key first applications of these results, namely the new compact determinant representations for many observables such as scalar products and correlators.

\vfill
\noindent 

\end{quote}

\setcounter{footnote}{0}\renewcommand{\thefootnote}{\arabic{thefootnote}}

\end{titlepage}

\tableofcontents

\section{Introduction}

Integrable systems play a prominent role in theoretical physics, as they help us to develop new intuition about nontrivial phenomena by learning from cases that are exactly solvable. Models of this type, from integrable spin chains to solvable quantum field theories in 2d and more recently in 4d \cite{Beisert:2010jr}, have been intensively studied over many years. Yet frequently the computation of various observables presents a great challenge even despite the presence of integrable structures. This is especially true for `off-shell' quantities like correlation functions which remain to be fully understood in many important models.

One powerful approach to attack this challenge is based on the idea of separation of variables (SoV), pioneered largely by Sklyanin in the late 80's and early 90's \cite{Sklyanin:1984sb,Sklyanin:1987ih,Sklyanin:1991ss,Sklyanin:1992eu,Sklyanin:1992sm,Sklyanin:1995bm}. The idea of this method is to try and find variables in which dynamics of the model with many degrees of freedom reduces to a set of non-interacting one-dimensional systems. For instance, this is what famously happens with the hydrogen atom in spherical coordinates. In a certain sense, the SoV realises the very essence of the notion of integrability. It is generally expected that one should be able to implement this approach for any integrable system, including both classical and quantum models.  In the quantum case, which we will focus on in this article, the decoupling of different degrees of freedom then implies a great simplification of wavefunctions which factorise into 1d building blocks. Roughly speaking, they take the form
\beq
\label{pif}
    \Psi(x_1,\dots,x_L)\sim Q(x_1)\dots Q(x_L)
\eeq
where $x_i$ are the separated coordinates and $Q$ are typically the Baxter Q-functions well known in many integrability constructions. This simplification of the wavefunctions has many merits, in particular often leading to a solution for systems intractable using standard Bethe ansatz, as well as generically opening the way to computation of a wide range of nontrivial observables such as correlators. 

This program has been mainly explored for a range of models based on rank-1 $GL(2)$-type symmetry (including the celebrated $SU(2)$ Heisenberg XXX spin chain), where it has been very successful. It has led to numerous new results for spin chains and related models (see e.g. \cite{Niccoli:2011nj,Niccoli:2012ci,Grosjean:2012ay,Niccoli:2012vq,Niccoli:2012pi,Niccoli:2012mq,Faldella:2013qha,Faldella:2013qca,Kitanine:2014swa,Niccoli:2014sfa,Kitanine:2015jna,Levy-Bencheton:2015mia,Kitanine:2016pvg,Kitanine:2018gki,Niccoli:2020zla,Niccoli:2024usi}), including examples where traditional methods based on the Bethe ansatz fail  \cite{Derkachov:2001yn,Derkachov:2002wz,Derkachov:2002pb,Derkachov:2002tf,Derkachov:2003qb} (see \cite{Guica:2017mtd} for a recent application in gauge/gravity duality), 
and even in integrable 2d QFTs (see in particular \cite{Smirnov:1998kv,Negro:2013wga}). 

At the same time, although SoV is expected to be a very general method, its applications to higher rank models based on $GL(N)$-type symmetry have been rather limited for a long time. The foundations of SoV for quantum $GL(3)$ spin chains were proposed by Sklyanin already in the early 90's \cite{Sklyanin:1992sm}, followed by important insights for classical $GL(N)$ models \cite{Scott:1994dz,gekhtman1995separation} that were later extended to certain quantum $GL(N)$ systems \cite{smirnov2002separation}. Yet the range of explored models has been rather restricted. In particular, 
a concrete realisation of the whole SoV program for perhaps the most natural examples -- rational quantum $GL(N)$ spin chains -- has been lacking until recently.

The full implementation of SoV for $GL(N)$ spin chains was finally developed in recent years, and the aim of this  review is to describe these key results in a concise and accessible manner. In particular, two main questions were understood -- how precisely the factorisation of wavefunctions is realised, and how to use it for efficient computation of various  observables. For completeness, below we first very schematically review the key steps following their chronological development, before giving a more pedagogical presentation in the main text.

\paragraph{Brief overview of higher-rank SoV results.} The early papers of Sklyanin \cite{Sklyanin:1992eu,Sklyanin:1992sm} laid the foundation for SoV in models based on $GL(3)$ symmetry. Regarding the extension to $GL(N)$, the first main question was how the factorisation of wavefunctions is realised. Building on these earlier results, \cite{Gromov:2016itr} proposed the implementation of SoV and factorisation of wavefunctions for rational quantum $GL(N)$ spin chains with any $N$, along the way uncovering a new surprisingly compact construction for their eigenstates. That construction was later proven for $GL(3)$ in \cite{Liashyk:2018qfc} and was  also extended to the supersymmetric $GL(2|1)$ case in \cite{Gromov:2018cvh}. In these results a key role is played by generalisations of Sklyanin's B-operator to the $GL(N)$ case, found independently in three different but likely equivalent forms in \cite{Gromov:2016itr} and two earlier papers \cite{smirnov2002separation,Chervov:2007bb}.

Next, the authors of \cite{Maillet:2018bim} put forward a different approach to implementing SoV, based on explicitly building the SoV basis via the action of transfer matrices instead of using Sklyanin's B-operator. This led to a realisation of SoV (as well as the proof of the Bethe ansatz description of the spectrum) for higher rank $GL(N)$ spin chains \cite{Maillet:2018czd}. These ideas were then used in combination with new insights in \cite{Ryan:2018fyo} to obtain a proof of the earlier results from \cite{Gromov:2016itr}, as well as the realisation of SoV for more general representations at the spin chain sites. The approach of \cite{Maillet:2018bim} was then also extended to cases with $q$-deformations \cite{Maillet:2018rto}, other representations \cite{Maillet:2019nsy}, boundaries \cite{Maillet:2019hdq} and some supersymmetric cases \cite{Maillet:2019ayx}. Yet more general representations, along with further insights, were explored in \cite{Ryan:2020rfk}.

With factorisation of wavefunctions understood, the next important question was deriving the SoV measure, which plays a key role in computation of correlators. This was done for higher rank spin chains in \cite{Cavaglia:2019pow,Gromov:2019wmz}\footnote{see also \cite{smirnov2002affine,smirnov2002quantization,Martin:2015eea} for earlier explorations in the semiclassical approach} which presented the approach known as functional SoV (FSoV) and established new SoV results for a number of nontrivial scalar products and form factors in determinant form. Subsequently another approach to deriving the measure indirectly was proposed in \cite{Maillet:2020ykb}. The FSoV ideas were extended to more general representations and many other observables in \cite{Gromov:2020fwh}, and then in \cite{Gromov:2022waj} were remarkably utilised to find determinant results for all matrix elements for a complete set of operators.  Following this, extensions to boundary overlaps were explored in \cite{Ekhammar:2023iph}.


\paragraph{Related approaches and applications.}

Let us mention another important motivation for advancing SoV methods -- namely, the potential to apply them in the context of gauge/string holographic duality. Integrability shows up in the most famous example of holography which relates N=4 supersymmetric Yang-Mills theory (SYM) in 4d with string theory on ${\rm AdS}_5\times{\rm S}^5$, see \cite{Beisert:2010jr} for a review. This offers the remarkable prospect to solve, for the first time, a nontrivial 4d gauge theory (together with the dual string theory). Yet while the spectrum of N=4 SYM is by now under excellent control \cite{Gromov:2013pga}, the remaining challenge of computing correlators is much harder and still is not resolved satisfactorily despite impressive progress (see e.g. \cite{Basso:2015zoa,Bargheer:2017nne}). SoV offers a new and highly promising way to attack this problem and has already brought many fruitful results, starting first with remarkable structures uncovered in the semiclassical or perturbative regimes, see e.g. \cite{Vicedo:2008ryn,Kazama:2012is,Sobko:2013ema,Kazama:2014sxa,Jiang:2015lda,Kazama:2015iua,Kazama:2016cfl}. Together with massive simplifications observed in more recent SoV approaches in this  and related models \cite{Cavaglia:2018lxi,Giombi:2018qox,Cavaglia:2021mft,Bercini:2022jxo}\footnote{see also e.g. \cite{Derkachov:2018rot,Derkachov:2019tzo,Derkachov:2021rrf,Derkachov:2021ufp} for a related though in many ways different approach}, this lends further importance to understanding and extending SoV for a variety of models. Let us also note that SoV plays an important role within a pure mathematical context as well, being linked to deep ideas related to exploration of Hitchin systems, Gaudin models, $N=2$ 4d gauge theories and their dualities, mathematical structures of 2d CFTs and ultimately even the Langlands correspondence (see e.g. \cite{Frenkel:1995zp,Nekrasov:2009rc,Frenkel:2015rda,Teschner:2017djr,Jeong:2024onv}).

In this review we focus on recent progress in higher rank SoV for spin chains. We will try to avoid too many technicalities, which in any case can be found in the literature, and rather try to present the key ideas and insights in a pedagogical and slightly less formal way. The presentation is expected to be accessible to readers with a range of backgrounds as long as they are familiar with basic methods in quantum integrability such as the algebraic Bethe ansatz. 

For clarity, in this review we mainly focus on $SU(3)$ spin chains with fundamental representation at each site, sometimes also using the $SU(2)$ spin chain as an illustration. Most of the results we discuss can be extended to any $SU(N)$ as well as to other representations, and in some cases even to trigonometric models. We refer to the corresponding literature in different parts of the review. 

This article is structured as follows. In section \ref{sec:defs} we briefly review basic methods and objects (Q-functions, transfer matrices etc) used for integrable spin chains and introduce the notation we use. In section \ref{sec:b} we discuss the construction of the SoV basis in which wavefunctions factorise using Sklyanin's B-operator approach. In section \ref{sec:mn} we discuss an alternative construction of the SoV basis utilising the models' transfer matrices. In section \ref{sec:fsov} we introduce the functional SoV approach providing a shortcut to computing various overlaps in SoV. In section \ref{sec:dual} we explain how it is connected to the operator realisation in the spin chain Hilbert space via building the dual SoV basis. In section \ref{sec:corr} we outline how all these SoV methods can be used to obtain new determinant representations for a wide range of correlators and related observables. Lastly in section \ref{sec:concl} we outline future directions, while some technical details are presented in the appendix.

\section{The setup: $SU(N)$ spin chains and $Q$-functions}
\label{sec:defs}

In this section we present the notation we will use as well as discussing standard basic features of the spin chains we study, such as the Bethe ansatz solution. For pedagogical reviews on this, see the classic reference \cite{Faddeev:1996iy} and  more recent articles such as e.g. \cite{Slavnov:2019hdn,Slavnov:2018kfx,Levkovich-Maslyuk:2016kfv}.

The models we discuss are based on the rational R-matrix acting in $\mC^N\otimes \mC^N$,
\beq
	R_{12}(u)=(u-i/2) + iP_{12}
\eeq
where $P_{12}$ is the permutation operator and $u$ is a complex variable known as the spectral parameter. Since this R-matrix is $GL(N)$-symmetric we will usually call these models $GL(N)$ spin chains or often $SU(N)$ spin chains (as it is also common in the literature).

In this review we will discuss in detail only spin chains with fundamental representation of $GL(N)$  at each site. The space of states of the spin chain with $L$ sites (which we will sometimes call the Hilbert space) is then a tensor product of $L$ spaces
\beq
{\cal H}=\mC^N\otimes\mC^N\otimes\dots \otimes\mC^N\ .
\eeq
Next we introduce an auxiliary $\mC^N$ space, labelled by the index `0', and define the monodromy matrix as
\beq
\label{mond}
	T(u)=R_{01}(u-\theta_1)R_{02}(u-\theta_2)\dots R_{0L}(u-\theta_L) g\ .
\eeq
Here we have introduced inhomogeneities $\theta_\alpha$ at each site $\alpha=1,\dots,L$ and a twist 
\beq
	g={\rm diag}\(\lambda_1,\lambda_2,\dots,\lambda_N\)\ .
\eeq
These serve as deformations that make the model more general and also often help as regulators. Without loss of generality we assume that
\beq
    \lambda_1\lambda_2\dots\lambda_N=1
\eeq
Explicitly, $T(u)$ is an $N\times N$ matrix whose entries are operators on our Hilbert space,
\beq
\label{Tmat}
	T(u)=\begin{pmatrix}
		T_{11}(u)&\dots&T_{1N}(u)\\
		\vdots&\ddots&\vdots\\
		T_{N1}(u)&\dots&T_{NN}(u)
	\end{pmatrix}\ ,
\eeq
They satisfy a number of commutation relations (defining what is called a Yangian algebra) which follow from the RTT identity
\beq
\label{RTT}
	R_{ab}(u-v){T_a(u)} {T_b(v)}=
	{T_b(v)}{T_a(u)} R_{ab}(u-v)\ ,
\eeq
that itself is a consequence of the Yang-Baxter equation satisfied by the R-matrix.

The trace of the monodromy matrix is known as the transfer matrix,
\beq
	t_1(u)={\rm tr}_0 T(u)=\sum_{n=1}^N T_{nn}(u)\ , \
\eeq
and it defines a commutative family of operators
\beq
	\[t_1(u),t_1(v)\]=0,\
\eeq
in whose joint eigenstates $\ket{\Psi}$ we will be interested -- in particular, in physically important cases the Hamiltonian is expressed in terms of these operators. For higher rank spin chains beyond $SU(2)$ the usual transfer matrix is to be diagonalised together with $N-2$ higher transfer matrices that correspond to antisymmetric representations in the auxiliary space.

We will also use the 'vacuum state' defined as
\beq
\label{def0}
	\ket{0}=\begin{pmatrix}1\\0\\ \vdots \\ 0\end{pmatrix}\otimes\begin{pmatrix}1\\0\\ \vdots \\ 0\end{pmatrix}
	\otimes\dots \otimes \begin{pmatrix}1\\0\\ \vdots \\ 0\end{pmatrix}\ ,
\eeq
which is a trivial eigenstate of the model.

Lastly, we will use a shorthand notation for shifts of the spectral parameter:
\beq\label{defpm}
	f^\pm\equiv f(u\pm i/2)\ , \ f^{[+a]}\equiv f(u+ia/2)\ .
\eeq

In the next subsection we discuss in more detail the $SU(2)$ and $SU(3)$ cases. 

\subsection{$SU(2)$ case}

Let us give some details on the Bethe ansatz solution of the spin chain for the $SU(2)$ case. Each state corresponds to a set of Bethe roots $u_1,\dots,u_M$ that are fixed by Bethe ansatz equations
\beq
\label{su2bae}
	\prod_{n=1}^L\frac{u_j-\theta_n+i/2}{u_j-\theta_n-i/2}=\frac{\lambda_2}{\lambda_1}
	\prod_{k\neq j}^M\frac{u_j-u_k+i}{u_j-u_k-i}\ 
\eeq
(we note that here $\lambda_2=1/\lambda_1)$. They can also be nicely packaged into a Q-function,
\beq
    \label{qtw}
    Q_1(u)=\lambda_1^{iu}
    \prod_{k=1}^M(u-u_k)
\eeq
whose roots are thus the Bethe roots. We will call this type of expression, i.e. a polynomial times an exponential, a \textit{twisted polynomial}. The Bethe equations have a discrete set of solutions with $M=0,1,\dots,L$ which (in the twisted case we consider) are in 1-to-1 correspondence with eigenstates of the spin chain.

The transfer matrix eigenvalue in these terms can be written as
\beq
\label{Tsl2Q}
\tau_1 = {Q^+_{\theta}}
\frac{
Q_{1}^{--}
}{Q_{1}}
+
Q_{\theta}^-
\frac{
Q_1^{++}
}{Q_1}
\;
\eeq
where $Q_\theta$ encodes the inhomogeneities,
\beq
    Q_\theta(u)=\prod_{\alpha=1}^L(u-\theta_\alpha)
\eeq
The above expression for $\tau_1$ becomes polynomial when the Bethe equations are satisfied. These can be recast as
\beq\la{BAEgeneralSL2}
\frac{Q_\theta(u^1_k+i/2)}
{Q_\theta(u_k^1-i/2)}
=-\frac{Q_1(u_k+i)}{Q_1(u_k-i)}
\;\;,\;\;k=1,\dots,M\;.
\eeq

Instead of the Bethe equations, one can use \eq{Tsl2Q} written in the form known as the Baxter equation,
\beq
\label{sl2bax1}
{Q^+_{\theta}}
{
Q_{1}^{--}
}
+
Q_{\theta}^-
{
Q_1^{++}
}-\tau_1{Q_1}=0
\;.
\eeq
Taking this equation and requiring $Q_1$ to be a twisted polynomial of the form \eq{qtw} and $\tau_1$ to be a degree $L$ polynomial fixes both of them (up to a discrete set of solutions) and in particular fixes all the roots of $Q_1$. This provides an equivalent alternative way to find Bethe roots without invoking Bethe equations directly.

\subsection{$SU(3)$ case}

For $SU(3)$ spin chains the Bethe ansatz is more complicated. We will have a whole set of Q-functions, out of which we will use here only three: $Q_1, Q_{12}, Q_{13}$  (we refer to \cite{Gromov:2017blm,Levkovich-Maslyuk:2019awk,Kazakov:2018ugh} for a recent review of Q-systems in general). The function $Q_1$ encodes the momentum-carrying Bethe roots, while the others contain two (dual) sets of auxiliary Bethe roots. All of these functions are again twisted polynomials and we will denote them as
\beq
    Q_1(u)=\lambda_1^{i u}\prod_{i=1}^{M_1}(u-u^{1}_k), \ Q_{12}(u)=(\lambda_1\lambda_2)^{i u}\prod_{i=1}^{M_{12}}(u-u^{12}_k), \ Q_{13}(u)=(\lambda_1\lambda_3)^{i u}\prod_{i=1}^{M_{13}}(u-u^{13}_k)
\eeq
The Bethe roots are fixed from the Bethe equations that now read
\beqa
\frac{Q_\theta(u_k^1+i/2)}
{Q_\theta(u_k^1-i/2)}
&=&-\frac{Q_1(u^1_k+i)}{Q_1(u^1_k-i)}
\frac{Q_{12}(u^1_k-\tfrac{i}{2})}{Q_{12}(u^1_k+\tfrac{i}{2})}\;\;,\;\;k=1,\dots,M_{1}\\
1&=&-
\frac{Q_{12}(u^{12}_k+i)}{Q_{12}(u^{12}_k-i)}
\frac{Q_{1}(u^{12}_k-\tfrac{i}{2})}{Q_{1}(u^{12}_k+\tfrac{i}{2})}\;\;,\;\;k=1,\dots,M_{12}\\
1&=&-
\frac{Q_{13}(u^{13}_k+i)}{Q_{12}(u^{13}_k-i)}
\frac{Q_{1}(u^{13}_k-\tfrac{i}{2})}{Q_{1}(u^{13}_k+\tfrac{i}{2})}\;\;,\;\;k=1,\dots,M_{13}
\eeqa
The Q-functions also satisfy a QQ-relation that reads here
\beq
    Q_{12}^+Q_{13}^- - Q_{12}^-Q_{13}^+\propto Q_1
\eeq
and which implies that their degrees are related by $M_1=M_{12}+M_{13}$. As we will see, Q-functions of different types play important roles in the SoV approach. 


For the $SU(3)$ case we have two transfer matrices, corresponding respectively to the  fundamental and antisymmetric irreps in the auxiliary space. Their eigenvalues can be written in terms of Q's as 
\beqa
\label{tau1Q}
\tau_1(u)&=&
Q_\theta(u+i/2)
\frac{Q_1^{--}}{Q_1}
+
Q_\theta(u-i/2)
\frac{Q_1^{++}}{Q_1}
\frac{Q_{12}^-}{Q_{12}^+}
+
Q_\theta(u-i/2)
\frac{Q_{12}^{[+3]}}{
Q_{12}^{+}
}\;,\\
\label{tau2Q}
\tau_2(u)&=&
Q_\theta(u+i/2)
\frac{Q_{12}^{[-3]}}{Q_{12}^{-}}
+
Q_\theta(u+i/2)
\frac{Q_1^{--}}{Q_1}
\frac{Q_{12}^+}{Q_{12}^-}
+
Q_\theta(u-i/2)
\frac{Q_{1}^{++}}{
Q_{1}
}\;.
\eeqa


Like for $SU(2)$, we can equivalently fix the Q-functions from the Baxter difference equations. Here they are 3rd order equations of the form
\beq
\label{Bax1sl3}
   {Q_\theta Q_\theta^{[-2]}}Q_1^{[+3]}
   - 
    \tau_{2}^{+}{Q_\theta^{[-2]}}Q_1^{+}
    +
     {Q_\theta^{[+2]}}\tau_1^{-}Q_1^{-}
      -
{Q_\theta^{[+2]}}Q_\theta Q_1^{[-3]}=0
\eeq
and
\beq
\label{Bax12s}
Q_\theta^{[+1]}Q_{1a}^{[-3]}-\tau_2Q_{1a}^- + \tau_1Q_{1a}^+ - Q_\theta^{[-1]}Q_{1a}^{[+3]}=0\;\;,\;\;a=2,3\;.
\eeq
For example, demanding all three solutions of the first equation  to be (twisted) polynomials together with polynomiality of $\tau_1,\tau_2$ fixes the set of solutions in 1-to-1 correspondence with states of the spin chain (equivalently to the Bethe ansatz). Then the remaining Q's can be found from the second Baxter equation. Other solution methods are also possible.

In the next section we proceed to discussing the fundamentals of the SoV construction, in particular the factorisation of wavefunctions for spin chains.

\section{The B-operator and SoV basis}
\label{sec:b}

\subsection{$SU(2)$ case}

For the $SU(2)$ spin chain the monodromy matrix has size $2\times 2$ and its elements are often denoted as
\beq
\label{tabcd}
    T(u)=\begin{pmatrix}
        A(u) & B(u) \\ C(u) & D(u)
    \end{pmatrix}
\eeq
so that $A(u),B(u),C(u)$ and $D(u)$ are operators on the Hilbert space of the spin chain. In this notation the tranfer matrix reads
\beq
    t_1(u)=A(u)+D(u)
\eeq
The operator $B(u)$ is also particularly important because it serves as a creation operator that builds eigenstates $\ket{\Psi}$ of the transfer matrix when acting on the vacuum state. That is, we have
\beq
\label{PB}
    \ket{\Psi}\propto B(u_1)\dots B(u_M)\ket{0}
\eeq
where $u_k$ are the Bethe roots satisfying the $SU(2)$ Bethe equations \eq{su2bae}. For the case with generic twist $\lambda_1$  and inhomogeneities $\theta_k$, the solutions to the Bethe equations with the number of roots $M\leq L$ are in 1-to-1 correspondence with the spin chain eigenstates and the construction \eq{PB} provides the full set of eigenstates (see e.g. \cite{Chernyak:2020lgw} and references therein).

It turns out that the same B-operator plays a central role in the SoV construction. We discuss this in the next subsection.

\subsubsection{General SoV construction for $SU(2)$-type models}
\label{sec:su2g}

The general SoV construction for the $SU(2)$ case was described in \cite{Sklyanin:1991ss, Sklyanin:1995bm}. Let us outline the main steps in this approach.


The key idea is that the eigenbasis of the $B$-operator turns out to be the desired SoV basis in which the wavefunctions factorise. More concretely, one assumes that $B(u)$ is a polynomial of degree $L$. This is not actually true for the spin chains described above (where $B$ has degree $L-1$), but it and other properties used below become true after some technical improvements we discuss in the next subsection, so it is useful to proceed here with this in mind. Since
\beq
    [B(u),B(v)]=0
\eeq
its coefficients form a commutative family and we can define its zeros as operators $x_k$,
\beq
    B(u)=B_0\prod_{\alpha=1}^L(u-x_\alpha)
\eeq
where the leading coefficient $B_0$ is also an operator.
One also assumes that we can diagonalize all the $x_\alpha$ operators simultaneously (which is generically possible since they commute) and we label their common eigenbasis by their eigenvalues, $\bra{\svx}=\bra{\svx_1,\dots,\svx_L}$, so\footnote{Note that we should distinguish the operators $x_k$ and their eigenvalues $\svx_k$, here we denote them with different fonts.}
\beq
    \bra{\svx}B(u)=B_0\prod_{\alpha=1}^L(u-\svx_\alpha)
\eeq
Then, if for the model in question the eigenstates can be built in the form \eq{PB}, we see that
\beq
    \bra{\svx}\Psi\rangle \propto \prod_{j=1}^M\prod_{\alpha=1}^L (u_j-\svx_\alpha)
\eeq
This is almost the product of Q's, up to twist factors which however depend only on the values of $\svx$'s and not on $\ket{\Psi}$. The normalisation of the SoV basis and the eigenstates $\ket{\Psi}$ can be fixed such that (see e.g. \cite{Gromov:2020fwh} for details)
\beq\label{psif1}
    \bra{\svx}\Psi\rangle=\prod_{\alpha=1}^LQ_1(\svx_\alpha)
\eeq
Thus we have achieved precisely the factorisation of the wavefunction in the basis $\bra{\svx}$ which can thus be rightly called the SoV basis. The form of \eq{psif1} is exactly as advertised in the Introduction in \eq{pif}.

We see that the operators $x_k$ play the role of the separated \textit{coordinates} as their eigenvalues appear in the factorised wavefunction. One can also define the corresponding conjugated \textit{momenta} using now the operators $A$ and $D$. For reasons of space we do not discuss this in detail here, rather we refer to \cite{Sklyanin:1991ss,Sklyanin:1995bm} for details of this construction (see also e.g. \cite{Gromov:2016itr} for details on its practical realisation for the spin chain).

Let us mention that the factorisation of wavefunctions in the eigenbasis of $B$ is typically still true even if the eigenstates construction of the kind \eq{PB} is \textit{not} available. This arises e.g. in infinite-dimensional representations without highest or lowest weight such as the principal series irreps (see e.g. \cite{Derkachov:2001yn}). While we will not discuss details here, this is in fact one of the important motivations for developing the SoV approach which can thus work even in cases when traditional Bethe Ansatz fails. 

In the next subsection we will discuss how this general SoV construction is implemented for our $SU(2)$ spin chain.

\subsubsection{Realization for $SU(2)$ spin chains}

If we try to directly realise the SoV construction from the previous subsection for our $SU(2)$ spin chain, then despite the simplicity of the model we immediately run into a rather inconvenient technical problem. Namely, the usual $B$-operator is not diagonalisable (its normal form is a Jordan cell with eigenvalues equal to zero). The reason for that is that it is a nilpotent operator, since it lowers the eigenvalue of the total spin operator, so acting multiple times on any state we will eventually flip all the spins and reach the state with all spins down which is annihilated by $B$. 

There are several ways to overcome this difficulty by introducing some kind of further deformation. Here we will follow the approach of \cite{Gromov:2016itr} (see also e.g. \cite{Kazama:2013rya} for a slightly different technique). Namely, we will define what was called in \cite{Gromov:2016itr} the 'good' monodromy matrix by introducing an extra similarity transformation in the auxiliary space, i.e.
\beq
    T^{\g}(u)=K^{-1}T(u)K
\eeq
where $K$ is a constant ($u$-independent) matrix acting only in the auxiliary space. We will assume $K$ to be in generic position (some  particularly simple possible choices of $K$ are given in \cite{Gromov:2016itr}). Accordingly we define the 'good' $A,B,C,D$ operators as the entries of $T^{\g}$,
\beq
    T^{\g}(u)=\begin{pmatrix}
        A^{\g}(u) & B^{\g}(u) \\ C^{\g}(u) & D^{\g}(u)
    \end{pmatrix}
\eeq
The 'good' operators are simply linear combinations of the original $A,B,C,D$. Notice also that the trace of the monodromy matrix is unaffected by this procedure, so we are still diagonalising the same set of conserved charges. Furthermore, since the R-matrix is $GL(2)$-symmetric, one can straightforwardly show that as a consequence of the RTT relation they satisfy the same commutation relations as the original $A,B,C,D$. In particular, the $B^{\g}$ operators still form a commutative family,
\beq
    [B^{\g}(u),B^{\g}(v)]=0
\eeq

Now, the whole point of the construction is that unlike the original $B$ the new operator $B^{\g}$ is no longer nilpotent and can actually be diagonalised. At the same time, it turns out that it retains the property of being a creation operator as it still generates the states by repeated action on the vacuum as in \eq{PB}:
\beq
    \ket{\Psi}=B^{\g}(u_1)\dots B^{\g}(u_M)\ket{0}
\eeq
where $u_k$ are the same Bethe roots as before. This property is not that obvious and is actually quite surprising, since $B^{\g}$ is a linear combination of all four $A,B,C,D$ operators, but it can be proven rigorously \cite{sklyanin1989new,Belliard:2018pie} (it is also related to the approach known as Modified Algebraic Bethe Ansatz \cite{Belliard:2018pvg}). 

As a result, using $B^{\g}$ in place of $B$ we can realise all the general steps from section \ref{sec:su2g} and build the SoV basis (the eigenbasis of $B^{\g}$) in which the transfer matrix eigenstates will factorise according to \eq{psif1}. One also finds that the eigenvalues of the separated variables $x_k$ are given by\footnote{Strictly speaking, only symmetric combinations of the $x_k$ are initially well defined as they appear as coefficients of powers of $u$ in a Taylor expansion of the operator $B(u)$. But then for each eigenstate of $B(u)$ we can define individual $x_k$ eigenvalues as a particular ordering of the zeros of the eigenvalue of $B(u)$ on that state. We find that we can do this in such a way that the spectrum of individual $x$'s is given by \eq{xsu2}.}
\beq
\label{xsu2}
    \svx_\alpha=\theta_\alpha\pm i/2
\eeq
We see that each variable $x_\alpha$ is associated with one of the sites of the spin chain. Notice that we have $L$ separated variables, and the eigenbasis $\bra{\svx}$ is labelled by all possible $2^L$ combinations of their eigenvalues \eq{xsu2}, thus giving a complete basis in our $2^L$-dimensional Hilbert space. We will not discuss rigorous proofs of these statements but refer to \cite{Sklyanin:1991ss,Sklyanin:1995bm,Maillet:2018bim}.

\subsection{$SU(3)$ case}

For the $SU(3)$ case the operator which should provide separated variables was proposed by Sklyanin in \cite{Sklyanin:1992sm}. Now it is not just a single entry of the monodromy matrix, like $B=T_{12}$ for $SU(2)$, but a polynomial of degree 3 in the entries $T_{ij}$ ($i,j=1,\dots,3$). To make its structure more clear it is useful to introduce \textit{quantum minors} of the monodromy matrix, which are defined much like the usual minors (determinants of submatrices), but with additional shifts of the spectral parameter:
\begin{align}\label{qminors}
    T\left[^{i_1 \dots i_a}_{j_1\dots j_a}\right](u) & =\sum_{\sigma}(-1)^{{\rm deg}\,\sigma\,}T_{i_{1}j_{\sigma(1)}}(u+i(a-1))T_{i_{2}j_{\sigma(2)}}(u+i(a-2))\dots T_{i_{a}j_{\sigma(a)}}(u) \\
    &= \sum_{\sigma}(-1)^{{\rm deg}\,\sigma\,}T_{i_{\sigma(1)}j_1}(u)T_{i_{\sigma(2)}j_2}(u+i)\dots T_{i_{\sigma(a)}j_a}(u+i(a-1))
\end{align}
Notice also that the entries $T_{ij}$ do not commute so their order is important here. Quantum minors in general have a variety of algebraic properties \cite{molev2007yangians}. Here using them we can write Sklyanin's $B$-operator for $SU(3)$ as
\begin{equation}\label{Bdef3}
    B=\sum_{j=1,2}T\left[^{j}_{3}\right]T^{[-2]}\left[^{12}_{j3}\right]
\end{equation}
Explicitly this gives
\beqa
\label{B3}
	B(u)&=&T_{23}(u)T_{12}(u-i)T_{23}(u)-T_{23}(u)T_{22}(u-i)T_{13}(u) \\ \nn
	&+&
	T_{13}(u)T_{11}(u-i)T_{23}(u)-T_{13}(u)T_{21}(u-i)T_{13}(u)
	\ .
\eeqa
Like for $SU(2)$, one can show \cite{Sklyanin:1992sm} that these operators form a commutative family, 
\beq
    [B(u),B(v)]=0
\eeq
Then for a general $SU(3)$-type model one expects that the $B$-operator is a polynomial of degree $3L$ whose zeros provide the separated coordinates,
\beq
    B(u)=B_0\prod_{\alpha=1}^{L}\prod_{a=1}^3(u-x_{\alpha,a})
\eeq
Accordingly the eigenbasis of $B$ in general should be the SoV basis.

\subsubsection{Improved operator and construction of eigenstates}
\label{sec:tg3}

In practice, for the $SU(3)$ spin chain we run into a difficulty similar to that for $SU(2)$ -- namely, this $B$-operator is nilpotent and cannot be diagonalised. However, as described in detail in \cite{Gromov:2016itr}, we can deal with this issue in the same way as before by introducing an extra similarity transformation in the auxiliary space and defining the improved monodromy matrix as 
\beq
\label{tgd}
    T^{\g}(u)=K^{-1}T(u)K
\eeq
where $K$ now is a constant $3\times 3$ matrix. Then we build the improved operator $B^{\g}$ out of the entries of the new monodromy matrix,
\beq\label{bgd}
    B^{\g}=\left. B\right|_{T_{ij}\to T_{ij}^{\g}}
\eeq
Notice that $T_{ij}^{\g}$ satisfy the same commutation relations as the original $T_{ij}$, which in particular ensures commutativity of the new $B$-operators,
\beq
    [B^{\g}(u),B^{\g}(v)]=0
\eeq

Now we can implement the whole Sklyanin's program as $B^{\g}$ is diagonalisable. Furthermore, remarkably, we observe that this operator allows one to build eigenstates just as for the $SU(2)$ case, simply by repeated action on the vacuum,
\beq
\label{psu3}
    \ket{\Psi}\propto B^{\g}(u^1_1)\dots B^{\g}(u^1_M)\ket{0}
\eeq
Here $u_k^1$ are the roots of $Q_1$, or in other words the momentum-carrying Bethe roots. The construction \eq{psu3} is highly surprising as it has a much simpler structure than the standard nested Bethe ansatz. The latter involves first constructing the wavefunctions of an auxiliary $SU(2)$ spin chain and then using them as coefficients in a nontrivial polynomial combination of matrix elements $T_{ij}$ (involving exponentially many terms, of order $2^M$ where $M$ is the number of Bethe roots) which acts on the vacuum. Here instead we have no need for any recursion and the complexity in $M$ is just linear. Notice also that there is even no need to invoke the auxiliary Bethe roots at all at this stage -- instead we can find $Q_1$ (and its roots that feature in the construction) from the 3rd order Baxter equation \eq{Bax1sl3} by requiring the polynomiality of all three of its solutions as well as the transfer matrices $\tau_1$ and $\tau_2$. This fixes a discrete set of solutions which are in 1-to-1 correspondence with spin chain eigenstates.  

Despite its compact and rather natural form, the construction \eq{psu3} is highly nontrivial to prove. Initially, for several special cases it was proven in \cite{Gromov:2016itr}. Then it was proven in general for $SU(3)$ in a rather nontrivial fashion in \cite{Liashyk:2018qfc}, and finally in \cite{Ryan:2018fyo} its algebraic structure was significantly clarified and a different proof was provided, valid in fact for any $SU(N)$. Let us also mention that this compact construction is not expected to work for spin chains with arbitrary representations of $GL(N)$ on the sites, but rather for a large class of irreps (roughly speaking with only one nonvanishing Dynkin label) including in particular the fundamental representation (see \cite{Liashyk:2018qfc,Ryan:2018fyo} for details).

\subsubsection{SoV realisation for $SU(3)$}

Having the $B^{\g}$ operator one can now diagnolise it and construct the SoV basis as its eigenbasis. We find that a trivial scalar factor of degree $L$ can be removed, leaving a polynomial of degree $2L$ of the form 
\beq
    B(u)=B_0Q_\theta^{[-3]}\prod_{\alpha=1}^{L}(u-x_{\alpha,1})(u-x_{\alpha,2})
\eeq
That is, now for each site $\alpha$ of the spin chain we have not one but two separated variables $x_{\alpha,1}, x_{\alpha,2}$. We denote the eigenbasis which diagonalises them as $\bra{\svx}$ so (with appropriate normalisations of the states) \footnote{Recall that the $B$-operators form a commutative family so the $x$-operators can be all simultaneously diagonalised.}
\beq
    \bra{\svx}\Psi\rangle=\prod_{\alpha=1}^{L}\prod_{a=1}^2Q_1(\svx_{\alpha,a})
\eeq
Furthermore, one finds \cite{Gromov:2016itr} that at any site $\alpha$ there are 3 possibilities for the values of the corresponding SoV coordinates:
\beq    (\svx_{\alpha,1},\svx_{\alpha,2})=\{\(\theta_\alpha-\tfrac i2,\theta_\alpha-\tfrac i2\),\(\theta_\alpha+\tfrac i2,\theta_\alpha-\tfrac i2\),\(\theta_\alpha+\tfrac i2,\theta_\alpha+\tfrac i2\)\}
\eeq
This can be summarised as
\beq
    \svx_{\alpha,a}=\theta_\alpha-i/2+in_{\alpha,a}
\eeq
with integers $n_{\alpha,a}$ satisfying
\beq
    0\leq n_{\alpha,2}\leq n_{\alpha,1}\leq 1
\eeq
and as result we see that the SoV basis contains $3^L$ elements, exactly matching the dimension of the Hilbert space.

\subsection{Extension to any $SU(N)$}
\label{sec:sun}

The results described above for $SU(3)$ can be extended in fact to any $SU(N)$. Let us briefly comment on this here.

Following the classical case \cite{Scott:1994dz,gekhtman1995separation}, the B-operator for the quantum spin chains for any $SU(N)$ was proposed in \cite{smirnov2002separation,Chervov:2007bb, Gromov:2016itr} in three rather different forms. Their equivalence is expected but remains to be verified and proven. The one from \cite{Gromov:2016itr} has perhaps the most explicit form and is given in appendix \ref{app:bc}. Then, using 'good' versions of the operators like for $SU(2)$ and $SU(3)$ above, one finds \cite{Gromov:2016itr} that this $B$-operator again provides the SoV basis as well as leading to the compact construction of eigenstates exactly as for $SU(3)$ above in \eq{psu3}. Remarkably, that construction  has the same compact form for any $SU(N)$, again without any recursion. These statements were proven in \cite{Ryan:2018fyo}\footnote{Strictly speaking the proof was not spelled out for the most general K-matrix, but it is likely possible to extend it to this case.} for all values of $N$.

In the next section we discuss another approach to implementing SoV which avoids the use of the $B$-operator.

\section{The SoV basis from action of T-operators}
\label{sec:mn}

In this section we discuss another way of  constructing the SoV basis that was proposed in \cite{Maillet:2018bim,Maillet:2018czd}. To keep this review relatively brief, we only outline the main features of this remarkable approach here, and refer the reader to these papers for further details and  rigorous derivations of many important statements.

The main idea is the very general observation that action of any conserved charges, corresponding to operators $\widehat I_\alpha$, on a generically chosen vector $\bra{\Omega}$ (not an eigenvector of the charges) leads to a factorized representation for the wavefunctions. Concretely, taking $L$ conserved charges $\widehat I_1,\widehat I_2,\dots \widehat I_L$ and constructing a bra vector
\beq
    \bra{X}=\bra{\Omega}\widehat I_1\widehat I_2\dots \widehat I_L
\eeq
allows one to view this vector as part of an SoV-like basis since its scalar product with the eigenstate of the model in question factorizes (up to a normalisation $\bra{\Omega}\Psi\rangle$):
\beq
    \bra{X}\Psi\rangle=\bra{\Omega}\widehat I_1\widehat I_2\dots \widehat I_L\ket{\Psi}=\bra{\Omega}\Psi\rangle \prod_{\alpha=1}^LI_\alpha
\eeq
Here we simply acted with the $\widehat I_\alpha$ operators onto their eigenstate $\ket{\Psi}$, producing their eigenvalues. Of course, in practice various nontrivial questions arise, such as independence of these states and completeness of the resulting basis, yet the idea behind the approach is rather general. 

For the spin chain models, one useful way to implement this construction is to use as $\widehat I_\alpha$ the transfer matrices (e.g. the simplest one, $t_1(u)$) evaluated at special values of the spectral parameter \cite{Maillet:2018bim,Maillet:2018czd}. For instance, in the $SU(2)$ case from the form of the transfer matrix \eq{Tsl2Q} we see that when it is evaluated at $u=\theta_\alpha-i/2$, the first term will vanish and we get\footnote{To be precise, recall that we denote the transfer matrix by $t_1(u)$ while \eq{Tsl2Q}  contains its eigenvalue denoted by $\tau_1(u)$. The Q-functions appearing in the rhs of \eq{tcr} should be understood as the Q-operators, which are defined by having Q-functions as their eigenvalues.}
\beq\label{tcr}
    \frac{ t_1(\theta_\alpha-i/2)}{Q_\theta(\theta_\alpha-i)}=\frac{Q_1(\theta_\alpha+i/2)}{Q_1(\theta_\alpha-i/2)}
\eeq
We see that the operator in the l.h.s. replaces a Q-function evaluated at one value of the separated coordinate (recall their eigenvalues from \eq{xsu2}) by the Q-function at the other value. Thus, for instance, taking the SoV basis state with values of the $x$-variables equal to $\svx_\alpha=\theta_\alpha-i/2$ as the starting state $\bra{\Omega}$, we can generate the rest of the SoV basis from the previous section by repeated action of \eq{tcr}. E.g. the state
\beq
    \bra{\Omega}\frac{ t_1(\theta_1-i/2)}{Q_\theta(\theta_1-i)}
\eeq
will have the eigenvalues of $x_\alpha$ operators equal\footnote{as is immediately found by taking its scalar product with $\ket{\Psi}$} to $\svx_1=\theta_1+i/2$ and all the other $\svx_{\alpha>2}=\theta_\alpha-i/2$.

The same idea can be used for higher rank models as well. For example, for $SU(3)$ from \eq{tau1Q} we see that at $u=\theta_\alpha+i/2$ the last two terms there will vanish and we will get a similar property to \eq{tcr}. Thus one can generate the SoV basis by using the two transfer natrices (in fundamental and antifundamental representation) in this way. This  provides the eigenbasis of the B-operator without in fact the need to consider this operator explicitly. We refer the reader to \cite{Maillet:2018bim,Maillet:2018czd} for full details of this construction. It has also been realised for other spin chains, including different representations \cite{Maillet:2019nsy}, boundary models \cite{Maillet:2019hdq}, trigonometric models \cite{Maillet:2018rto} and partially for supersymmetric spin chains \cite{Maillet:2019ayx}. One may hope that exploring the full scope of this approach in the future will lead to further applications.

Let us mention that in general an important conceptual step in implementing this approach is establishing the closure of the algebra of creation operators that construct the SoV basis (like e.g. \eq{tcr}). This issue has been settled for $SU(N)$-type models by virtue of fusion relations between transfer matrices. For the supersymmetric case this question becomes more nontrivial \cite{Maillet:2019ayx}.


The fact that the construction of \cite{Maillet:2018bim,Maillet:2018czd} diagonalises the B-operator was rigorously established in \cite{Ryan:2018fyo}. Thus one can build the SoV basis equivalently as an eigenbasis of B or by directly constructing it with repeated action of transfer matrices. Both points of view come useful in different situations, and as we will see it is often the combination of the two approaches that leads to fruitful results.

\section{Functional SoV and scalar products}
\label{sec:fsov}

Now that we have established the construction of the SoV basis, let us move on to a key application of SoV -- calculation of scalar products and correlators. This will be the subject of the rest of this review, and in this section we will introduce the approach to computing correlators known as functional SoV, mainly following \cite{Cavaglia:2019pow, Gromov:2019wmz, Gromov:2020fwh}. 

One of the main ingredients in calculation of correlators is known as the SoV measure. Let us first give an informal picture of it, starting with the example of the hydrogen atom. For that case, in the original variables $x,y,z$ the eigenstates are rather complicated, but in spherical coordinates $r,\theta,\phi$ the variables separate, so the wavefunctions greatly simplify and factorise as $\Psi=F_1(r)F_2(\theta)F_3(\phi)$. At the same time, to compute an overlap of two wavefunctions in the separated coordinates we we need to take into account a nontrivial measure, i.e. the volume element $r^2\sin\theta$ which is the Jacobian arising when we pass to the spherical coordinates,
\beq
    \int dx dy dz\to \int r^2\sin\theta dr d\theta d\phi\; 
\eeq
For more general integrable models, it is the analog of this Jacobian that is known as the SoV measure.


Let us explain this more concretely for spin chains. For simplicity let us first consider an overlap between a left and right eigenstate of the transfer matrix. Although these are of course zero for states corresponding to different eigenvalues, these observables are the starting point for more complicated calculations and turn out to capture a lot of key structures in SoV. Let us also note that we will not assume that left and right eigenstates are Hermitian conjugate (although this may be the case in many situations), rather we consider them simply e.g. as column and row vectors. Now, for $SU(2)$-type models one can typically construct both a right and a left SoV basis, in which the wavefunctions factorise,
\beq
\label{psif2}
    \bra{\svx}\Psi\rangle\sim Q(\svx_1)\dots Q(\svx_L) \ , \ \ \bra{\Psi}\svx\rangle\sim Q(\svx_1)\dots Q(\svx_L)
\eeq
and then write the resolution of identity in the form
\beq
\label{im1}
   1=\sum_\svx M_\svx \ket{\svx}\bra{\svx}
\eeq
Here the coefficient $M_\svx=(\bra{\svx}\svx\rangle)^{-1}$ is the SoV measure (see e.g. \eq{mxsu2} for an example for the $SU(2)$ case). Then inserting it into the scalar product between two states $A$ and $B$ that factorise as in \eq{psif2} we find that it becomes 
\beq
\label{fsp}
    \bra{\Psi_B}\Psi_A\rangle \sim \sum_\svx M_\svx \prod_{\alpha=1}^LQ_A(\svx_\alpha)Q_B(\svx_\alpha)
\eeq
As a result, the scalar product is written now in terms of Q-functions. In some models the separated variables may have a continuous spectrum and we will have an integral instead of a sum in \eq{fsp}. Similar formulas in terms of Q-functions can be derived for more complicated observables such as (at least some class of) correlation functions, opening the way to explore many of their nontrivial features. 


However, for higher rank models the same logic does not immediately give a useful answer. The reason is that the overlaps between left and right SoV states become much more complicated and no concise formulas are known for them. Instead, as we will see, one way to get a compact result is to factorise the left and right eigenstates in two \textit{different} SoV bases. As an outcome we will still find a formula of the type \eq{fsp} for various overlaps and correlators.

In fact, it turns out that there is a shortcut to deriving the formula for the overlap in terms of Q's, bypassing many technical difficulties such as the construction of the SoV bases themselves. This idea is known as functional SoV (FSoV) and in this section we will describe how it works. Then in the next section \ref{sec:dual} we will discuss its relation with the SoV basis and the rest of the operatorial SoV construction.

\subsection{Functional SoV for the $SU(2)$ case}

Let us start by illustrating the main ideas of functional SoV for the simplest $SU(2)$ case, in a way that will generalise to higher rank models. The main observation is that, since the eigenstates corresponding to different eigenvalues are orthogonal, and furthermore in the SoV basis they are given by Q-functions, there should be some kind of orthogonality property for Q-functions themselves. And as the Q-functions can be fixed solely from the Baxter equation, this orthogonality must be visible from this equation alone. 

To realise this idea, let us write the Baxter equation \eq{sl2bax1} in operator form as
\beq
	\widehat O Q_1=0 \ , \ \ \ \widehat O={Q_\theta^+}D^{-2}-{\tau_1}+{Q_\theta^-}D^{2}
\eeq
where $D$ is a shift operator,
\beq
    D f(u)=f(u+i/2)
\eeq
In this notation  $\widehat O$ is a 2nd order difference operator that annihilates the Q-function. Now we define the bracket
\beq
\label{br1}
	\langle f \rangle=\oint du \; \mu\; f
\eeq
where the measure is given by
\beq
	\mu=\frac{1}{Q_\theta^+Q_\theta^-} 
\eeq
and integral is around a big circle going around all poles of the integrand. 
The key fact is that the Baxter operator $\widehat O$ is \textit{self-adjoint} with respect to this bracket, i.e.
\beq
\label{sa1}
    \langle f\widehat Og\rangle=\langle g\widehat Of\rangle
\eeq
where $f$ and $g$ are arbitrary twisted polynomials (of the form \eq{qtw}). This property will lead to orthogonality by the arguments similar to usual quantum mechanics as we will see below. To prove this property we rewrite the l.h.s. of \eq{sa1} as 
\beq
    \langle f\widehat Og\rangle=\oint  \left(\frac{fg^{--}}{Q_\theta^-}-\frac{fg}{Q_\theta^+Q_\theta^-}+\frac{fg^{++}}{Q_\theta^+}\right)
    =\oint \left(\frac{f^{++}g}{Q_\theta^+}-\frac{fg}{Q_\theta^+Q_\theta^-}+\frac{f^{--}g}{Q_\theta^-}\right)
\eeq
To obtain the last equality, we shifted the integration contour as $u\to u+i$ in the first term and as $u\to u-i$ in the 3rd term -- the goal being to move the shifts of the argument from $g$ onto $f$. In the r.h.s. of this equation we now recognise precisely $\langle g\widehat Of\rangle$, so we conclude that \eq{sa1} holds.

Note that the origin of this property lies in the symmetric shifts by $\pm i/2$ in the coefficients of $Q_1^{++}$ and $Q_1^{--}$ in the Baxter equation -- this is what led to having $\widehat Of$ when starting from $\widehat Og$ and shifting the contours. In more general situations where the coefficients are different, requiring the self-adjointness property to hold typically leads to functional equations for $\mu$ that fix it to be a more involved function than here (see \cite{Gromov:2020fwh} for examples).

In fact the above derivation remains true if we multiply $\mu$ by any regular $i$-periodic function. We will see that it useful to define a family of $L$ measures as 
\beq\label{ma}
	\mu_\alpha=\frac{P_\alpha}{Q_\theta^+Q_\theta^-} \ \ , \ \alpha=1,\dots,L
\eeq
where the $i$-periodic numerator is given by
\beq
P_\alpha=\prod_{\beta\neq\alpha}^L(e^{2\pi u}-e^{2\pi(\theta_\beta+i/2)})
\eeq
The role of the numerator is to cancel all poles except for the ones at $u=\theta_\alpha\pm i/2$. Accordingly we define $L$ brackets as
\beq
	\langle f \rangle_\alpha=\oint \mu_\alpha\; f
\eeq
and we have
\beq
    \langle f\widehat Og\rangle_\alpha=\langle g\widehat Of\rangle_\alpha
\eeq

To derive orthogonality for Q's we now invoke an argument rather similar to standard quantum mechanics. Consider two states $A$ and $B$, so that
\beq
    \widehat O^AQ_1^A=0, \ \ \widehat O^BQ_1^B=0
\eeq
Note that the Baxter operator $\widehat O$ itself depends on the state as it contains the transfer matrix eigenvalue. Then using the self-adjointness property \eq{sa1} we have
\beq
\label{amb}
    \langle Q_1^B(\widehat O^A-\widehat O^B)Q_1^A\rangle_\alpha=\langle Q_1^B\widehat O^AQ_1^A\rangle_\alpha-\langle Q_1^B\widehat O^BQ_1^A\rangle_\alpha =0
\eeq
where in the last term we used \eq{sa1} to act with the $\widehat O^B$ operator to the left, onto $Q^B$ which is annihilated by it. Next, to write explicitly the l.h.s. of \eq{amb} let us write the transfer matrix eigenvalue as
\beq
    \tau_1^A=2\cos\phi u^L+\sum_{\alpha=1}^{L}I^A_{\alpha-1}u^{\alpha-1}
\eeq
where $I_\alpha$ are the eigenvalues of the coefficients of the transfer matrix (which mutually commute and play the role of the integrals of motion). Plugging this into \eq{amb} we get
\beq
\label{isys}
    \sum_{\beta=1}^{L}(I^A_{\beta-1}-I^B_{\beta-1})\langle Q_1^AQ_1^Bu^{\beta-1}\rangle_\alpha=0 
\eeq
For the twisted spin chain that we consider, the set of these integrals of motion uniquely identifies the state. Thus, if the two states $A$ and $B$ are different, then at least for some value of $\alpha$ we must have a nonvanishing difference $I_\alpha^A-I_\alpha^B\neq 0$. At the same time, \eq{isys} is a homogenous linear system for these differences which thus must have a nonzero solution. So its determinant must vanish, and we conclude that
\beq
\label{detsu2}
    \det \left|\langle Q_1^AQ_1^Bu^{\beta-1}\rangle_\alpha\right|_{\alpha,\beta=1,\dots,L}\propto \delta_{AB}
\eeq
As an example, for $L=2$ this determinant has the form
\beq
\begin{vmatrix}
	\langle {uQ_1^AQ_1^{B}}\rangle_1 & \langle {Q_1^AQ_1^{B}}\rangle_1 \\
	\langle {uQ_1^AQ_1^{B}}\rangle_2 & \langle {Q_1^AQ_1^{B}}\rangle_2 
	\end{vmatrix}
\eeq
We see that the determinant in the l.h.s. of \eq{detsu2} has precisely the property of the scalar product between states $A$ and $B$ as it vanishes for different states. One can also check that it is nonzero for identical states. This strongly suggests to identify it with the expected scalar product of the form \eq{fsp} in the SoV representation. In fact, expanding the determinant in \eq{detsu2} and evaluating each bracket by picking the poles at $\theta_\alpha\pm i/2$, we find that
\beq
    \det \left|\langle Q_1^AQ_1^Bu^{\beta-1}\rangle_\alpha\right|_{\alpha,\beta=1,\dots,L}=\sum_{\svx_\alpha=\theta_\alpha\pm i/2}C_{\svx_1,\dots,\svx_L}\prod_{\alpha=1}^LQ_1^A(\svx_\alpha)\prod_{\alpha=1}^LQ_1^B(\svx_\alpha)
\eeq
where the sum is over the $2^n$ possible combinations of the values of the variables $\svx_\alpha=\theta_\alpha\pm i/2$. This is precisely an expression of the form \eq{fsp}, where the Q-functions are evaluated right at the eigenvalues of the separated coordinates we discussed in \eq{xsu2} in section \ref{sec:b}. And computing the residues we find that  the coefficients $C_{\svx_1,\dots,\svx_L}$ reproduce\footnote{up to an irrelevant overall normalisation} the known SoV measure for $SU(2)$ (given explicitly in \eq{mxsu2} below).

The approach we described is often called functional separation of variables (or FSoV). We see that it provides a relatively simple shortcut to deriving the SoV measure using just the Baxter equation. In the next subsection we will see how to generalise this idea to $SU(3)$ and higher ranks.

\subsection{$SU(3)$ case}

Let us see now how to implement the FSoV approach now for the $SU(3)$ spin chain. Here we have two Baxter equations \eq{Bax1sl3}, \eq{Bax12s}, corresponding to two difference operators that we denote by $\widehat O$ and $\widehat O^\dagger$,
\beq
   \widehat OQ_1=0, \ \ \widehat O=\frac{1}{Q_\theta^{[+2]}}D^3
   - 
    \tau_{2}^{+}\frac{1}{Q_\theta Q_\theta^{[+2]}}D
    +
     \tau_1^{-}\frac{1}{Q_\theta Q_\theta^{[-2]}}D^{-1}
      -
\frac{1}{Q_\theta^{[-2]}}D^{-3}
\eeq
\beq
\label{od3}
	\widehat O^\dagger Q_{1,a+1}=0 \ , \ \ \ \widehat  O^\dagger={Q_\theta^+}D^{-3}-{\tau_2}D^{-1}+{\tau_1}D-{Q_\theta^-}D^{3}
\eeq
with $a=1,2$. It turns out that now instead of self-adjointness we find that these operators are adjoint to each other,
\beq
\label{oad}
    \langle f\widehat Og\rangle_\alpha=\langle g\widehat O^\dagger f\rangle_\alpha
\eeq
where the bracket is exactly the same as before, given by \eq{br1} with the measure \eq{ma}. This property can be proven just like in the $SU(2)$ case, by moving the contour in order to transfer the shifts of the argument from $g(u)$ onto $f(u)$. Now we use essentially the same trick as for $SU(2)$ to derive orthogonality. Consider two states $A$ and $B$ and the combination
\beq
\label{od2}
    \langle Q_1^A(\widehat O^{\dagger A}-\widehat O^{\dagger B})Q_{1,a+1}^B\rangle_\alpha=0
\eeq
which vanishes due to \eq{oad}. We can write the transfer matrices as
\beq
    \tau_a(u)=\chi_a(\lambda)u^L+\sum_{\alpha=1}^{L}I_{a,\alpha-1}u^{\alpha-1} \ , \ \ \ a=1,2
\eeq
where $I_{a,\alpha}$ are the eigenvalues of the integrals of motion and the leading term is the character defined by the twist,
\beq
    \chi_1(\lambda)=\lambda_1+\lambda_2+\lambda_3, \ \ \ \chi_2(\lambda)=\lambda_1\lambda_2+\lambda_1\lambda_3+\lambda_2\lambda_3
\eeq
Plugging this into \eq{od2} we find
\beq
    \label{su3lin}\sum_{b=1}^2\sum_{\beta=1}^L\langle Q_1^Au^{\beta-1}D^{3-2b}Q_{1,a+1}^B\rangle_\alpha\times (-1)^b(I_{b,\beta-1}^A-I_{b,\beta-1}^B)= 0\ , \ \ 
\eeq
with $a=1,2$ and $\alpha=1,\dots,L$. As before, the set of integrals of motion identifies the state uniquely, and here we have a linear system for their differences. So for $A\neq B$ its determinant must vanish and we get
\beq\label{detsl3}
\det_{(a,\alpha),(b,\beta)}\langle Q_1^A \;u^{\beta-1}\; {\cal D}^{3-2b} Q_{1,a+1}^B \rangle_\alpha\propto\delta_{AB}\;.
\eeq

Like for $SU(2)$, this determinant is thus a natural candidate for scalar product in SoV,
\beq
\label{pp3}
    \bra{\Psi_B}{\Psi_A}\rangle\propto \det_{(a,\alpha),(b,\beta)}\langle Q_1^A \;u^{\beta-1}\; {\cal D}^{3-2b} Q_{1,a+1}^B \rangle_\alpha
\eeq
Now it has a more complicated structure, but one can show, as we discuss in section \ref{sec:dual}, that it precisely corresponds to the wavefunction overlap evaluated in the appropriately chosen SoV basis.

For clarity, let us write out this determinant more explicitly, highlighting the two states ${\color{blue}A}$ and $\color{darkgreen}B$  with different colors. For $L=1$ it reads
\beq
\bra{{\color{darkgreen}\Psi_B}}{\color{blue}\Psi_A}\rangle\propto
	\begin{vmatrix}
	\langle {{\color{blue}Q_1^A}{\color{darkgreen}Q_{12}^{B+}}}\rangle_1 & \langle{{\color{blue}Q_1^A}{\color{darkgreen}Q_{12}^{B-}}}\rangle_1  \\
	\langle {{\color{blue}Q_1^A}{\color{darkgreen}Q_{13}^{B+}}} \rangle_1 & \langle{{\color{blue}Q_1^A}{\color{darkgreen}Q_{13}^{B-}}} \rangle_1
	\end{vmatrix}
\eeq
For $L=2$ we have
\beq
	\bra{{\color{darkgreen}\Psi_B}}{\color{blue}\Psi_A}\rangle\propto\begin{vmatrix}
	\langle {u{{\color{blue}Q_1^A}}{\color{darkgreen}Q_{12}^{B+}}}\rangle_1 & \langle {{{\color{blue}Q_1^A}}{\color{darkgreen}Q_{12}^{B+}}}\rangle_1 & 
    \langle{u{{\color{blue}Q_1^A}}{\color{darkgreen}Q_{12}^{B-}}} \rangle_1 & \langle {{{\color{blue}Q_1^A}}{\color{darkgreen}Q_{12}^{B-}}}\rangle_1  \\
	\langle {u{{\color{blue}Q_1^A}}{\color{darkgreen}Q_{12}^{B+}}}\rangle_2 & \langle {{{\color{blue}Q_1^A}}{\color{darkgreen}Q_{12}^{B+}}}\rangle_2 & \langle{u{{\color{blue}Q_1^A}}{\color{darkgreen}Q_{12}^{B-}}} \rangle_2 & \langle {{{\color{blue}Q_1^A}}{\color{darkgreen}Q_{12}^{B-}}}\rangle_2  \\
	\langle {u{{\color{blue}Q_1^A}}{\color{darkgreen}Q_{13}^{B+}}}\rangle_1 & \langle {{{\color{blue}Q_1^A}}{\color{darkgreen}Q_{13}^{B+}}}\rangle_1 & \langle{u{{\color{blue}Q_1^A}}{\color{darkgreen}Q_{13}^{B-}}} \rangle_1 & \langle {{{\color{blue}Q_1^A}}{\color{darkgreen}Q_{13}^{B-}}}\rangle_1  \\
	\langle {u{{\color{blue}Q_1^A}}{\color{darkgreen}Q_{13}^{B+}}}\rangle_2 & \langle {{{\color{blue}Q_1^A}}{\color{darkgreen}Q_{13}^{B+}}}\rangle_2 & \langle{u{{\color{blue}Q_1^A}}{\color{darkgreen}Q_{13}^{B-}}} \rangle_2 & \langle {{{\color{blue}Q_1^A}}{\color{darkgreen}Q_{13}^{B-}}}\rangle_2  \\
	\end{vmatrix}
\eeq
In terms of the structure, we see that this determinant has two main differences compared to $SU(2)$. First, the states $A$ and $B$ are represented by different types of Q-functions ($Q_1^A$ vs $Q_{12}^B,Q_{13}^B$). Second, there are shifts in the argument of the Q's (by $\pm i/2$) entering this expression. 

The same method allows one to generalise this result to any $SU(N)$. Let us highlight just one key feature here, namely that the Q-functions entering for state B will be (instead of $Q_{12}$ and $Q_{13}$ for $SU(3)$) the $N-1$ Q-functions that contain the Bethe roots at the deepest level of nesting, often denoted by $Q^{a}$ in the literature on Q-systems (with $a\neq 1$ in our case). We refer to \cite{Cavaglia:2019pow,Gromov:2019wmz,Gromov:2020fwh} for details. 


One can also generalise to an even a larger class of representations including in particular infinite-dimensional highest weight representations with arbitrary negative or even complex spin. In these latter cases especially proving the key conjugation property \eq{oad} is more tricky as the contours go along the real line and will cross some poles of the integrand when we shift them. Nevertheless all the relevant residues perfectly cancel, further indicating that the approach in question reveals a natural structure present in the model. For further discussion see \cite{Cavaglia:2019pow,Gromov:2020fwh}. For even more general representations such as the principal series one (without highest or lowest weight) the structure of the determinants that appear here seems to be more nontrivial and involves various ambiguities, but nevertheless many results in the FSoV approach were obtained for this case in \cite{Cavaglia:2021mft}.

Let us note that the structure of the result \eq{pp3} above and more generally the one for $SU(N)$ \cite{Cavaglia:2019pow, Gromov:2019wmz} is in line with the general expectations based on semiclassical arguments involving the quantization of the classical spectral curve \cite{smirnov2002quantization, smirnov2002affine}. In particular, the presence of shifts in arguments of Q's aligns very well with that structure as the shift operators correspond roughly speaking to the momenta conjugate to separated coordinates, and thus to one of the natural spectral curve parameters (see \cite{Cavaglia:2019pow} for further comments).


\section{Dual separated variables and SoV measure}
\label{sec:dual}

In the previous section we derived, purely from the Baxter equations, a determinant expression that serves as a candidate for overlaps of the wavefunctions in SoV. In this section we will link this expression to the operator realisation of SoV and argue that it indeed corresponds to scalar products of states in appropriately chosen SoV bases. We will also show how to extract from it the SoV measure explicitly. We will discuss only the $SU(3)$ case here and refer to \cite{Gromov:2019wmz,Gromov:2020fwh} for results for any $SU(N)$.

\subsection{Dual SoV basis and the SoV measure}

In the determinant formula \eq{detsl3} we derived above using FSoV, the Q-functions for states A and B enter in a different way: state A is represented by $Q_1$ whereas state B is represented by $Q_{12}$ and $Q_{13}$. This suggests the main idea exploited in \cite{Gromov:2019wmz} -- to try and factorise the left and right eigenstates (that is,  $\ket{\Psi}$ and $\bra{\Psi}$) in \textit{different} SoV bases. 

From section \ref{sec:b} we know the basis $\bra{\svx}$ in which the ket states $\ket{\Psi}$ factorise into a product of the functions $Q_1$. Remarkably, it was understood in \cite{Gromov:2019wmz} that one can construct a dual basis $\ket{\svy}$ in which the bra states $\bra{\Psi}$ factorise into components built from $Q_{12},Q_{13}$ that match the FSoV determinant formula \eq{pp3}. While the basis $\bra{\svx}$ diagonalises the B-operator $B^{\g}(u)$ from \eq{B3}, \eq{bgd}, it turns out that the dual basis $\ket{\svx}$ diagonalises a rather similar operator denoted as $C^{\g}(u)$. It differs from the B-operator only by the shifts in the arguments of $T_{ij}$ appearing in it\footnote{Notice that for the $SU(2)$ case the $C$-operator we discuss here is \textit{not} the $C$-operator of the $A,B,C,D$ family defined by \eq{tabcd}, rather it is the same $B$-operator again.},
\begin{equation}   C=\sum_{j=1}^2T\left[^{12}_{j3}\right]T\left[^{j}_{3}\right]\;
\end{equation}
in the same notation used in \eq{Bdef3}. And the 'good' version is defined by using an extra similarity transformation in the auxiliary space as before,
\beq
    C^{\g}=\left. C\right|_{T_{ij}\to T_{ij}^{\g}}
\eeq
where $T^{\g}$ is given by \eq{tgd}. 

Like the B-operator, the C-operator commutes with itself for different values of $u$,
\beq
    [C^{\g}(u),C^{\g}(v)]=0
\eeq
and (up to a trivial scalar factor that we will discard) is a polynomial of degree  $2L$ for $SU(3)$. Denoting its operator roots as $y_{\alpha,a}$ with $\alpha=1,\dots,L$ and $a=1,2$, we have in its eigenbasis
\beq
    C^{\g}(u)\ket{\svy}=C_0\prod_{\alpha=1}^L(u-y_{\alpha,1})(u-y_{\alpha,2})\ket{\svy}
\eeq
where $C_0$ is a trivial scalar prefactor. We will call the $y$-variables the \textit{dual} SoV coordinates, similarly to the roots of the $B$-operator\footnote{notice that $x$ and $y$ are not meant to be canonically conjugate in any sense, rather these are two sets of separated coordinates}. The spectrum of the $y$-variables is found to be 
\beq
    \svy_{\alpha,1}=\theta_\alpha-i/2+im_{\alpha,1}, \ \ \svy_{\alpha,2}=\theta_\alpha-3i/2+im_{\alpha,2}\ , \ \ \alpha=1,\dots, L
\eeq
with integer values for $0\leq m_{\alpha,2}\leq m_{\alpha,1}$.

Notice that the $C$-operator does not create spin chain eigenstates simply by acting on the vacuum in the style of \eq{PB}, unlike the B-operator. However, one can show that still the wavefunctions factorise in the eigenbasis of $C$, and the building blocks will be not single Q-functions but rather their antisymmetric combinations:
\beq
    \label{py}\bra{\Psi}y\rangle=\prod_{\alpha=1}^L(Q_{12}(y_{\alpha,1}+\tfrac i2)Q_{13}(y_{\alpha,2}+\tfrac i2)-(Q_{12}\leftrightarrow Q_{13}))
\eeq
To show this one invokes algebraic methods of \cite{Ryan:2018fyo}, see \cite{Ryan:2020rfk} for details. In fact, loosely speaking the $C$-operator can be understood as the realisation of $B$ for the conjugate representation, to which the bra eigenstates are naturally associated, and this explains the factorisation in \eq{py} observed initially for the conjugate irrep in \cite{Ryan:2018fyo}. 

Let us recall also that the usual SoV basis $\bra{\svx}$ can be built by repeated action of the transfer matrices on a reference state as discussed in section \ref{sec:mn} following \cite{Maillet:2018bim}. In fact the dual basis $\ket{\svy}$ can be constructed in a similar fashion as was understood in \cite{Gromov:2019wmz}. This construction often is very useful in order to establish various properties of the SoV bases and provides another  interesting realisation of the general ideas of \cite{Maillet:2018bim}.

The structure of \eq{py} is already highly similar to what one gets when evaluating the determinant \eq{pp3} by taking the integrals via residues. Showing a complete match requires further work and taking into account nontrivial cancellations when one expands the determinant. All of this was successfully worked out in \cite{Gromov:2019wmz,Gromov:2020fwh}. As a result, the final picture is the following. We consider a resolution of identity in the form
\beq
    \sum_{\svx, \svy}M_{\svy,\svx}
\ket{\svy}\bra{\svx}=1
\eeq
Here the SoV measure $M_{\svy,\svx}$ is the inverse of the matrix of overlaps $\langle\svx\ket{\svy}$. Next, we insert it into the overlap of two states $A$ and $B$, which gives
\beqa\la{oversov}
\langle \Psi_B|\Psi_A\rangle
&=&
\sum_{\svx, \svy}M_{\svy,\svx}\langle \Psi_B\ket{\svy}\bra{\svx}
\Psi_A\rangle\\
\label{osov2}
&=&\sum_{\svx, \svy}M_{\svy,\svx}
\prod_{\alpha=1}^L
{
\[Q^B_{12}(\svy_{\alpha,1}+\tfrac{i}{2})
Q^B_{13}(\svy_{\alpha,2}+\tfrac{i}{2})
-(Q^B_{12} \leftrightarrow Q^B_{13})\]}
{Q^A_1(\svx_{\alpha,1})
    Q^A_1(\svx_{\alpha,2})}\;.
\eeqa
Now we can compare this with the FSoV determinant  formula \eq{pp3} from section \ref{sec:fsov} and we find a perfect match! For example, taking some fixed length of the spin chain we can explicitly compute the overlaps matrix $\langle\svx\ket{\svy}$, compute its inverse $M_{\svy,\svx}$ and find that \eq{osov2} perfectly matches\footnote{It would also be interesting to prove this statement in full generality in the future. } the expansion of the determinant \eq{pp3}. This puts the determinant formula \eq{pp3} on a firm ground, providing its realisation at the level of states and operators in the spin chain Hilbert space and showing that it indeed gives the overlap of two eigenstates, factorised in the bases $\bra{\svx}$ and $\ket{\svy}$. 

A remarkable fact one can observe at this stage is that the overlaps matrix $M_{\svy,\svx}$ turns out to be independent of the twist, when the simple overall normalisations are chosen appropriately (we refer to \cite{Gromov:2020fwh} for details). We will discuss the reason for this below in subsection \ref{sec:comp}. 

Now, by expanding the FSoV determinants and carefully analysing their structure to account for all nontrivial cancellations, one can compare the result with \eq{osov2} and find the SoV measure $M_{\svy,\svx}$ in a completely explicit form -- another important outcome of the whole program.  Let us discuss briefly its structure. For $SU(2)$ the overlaps matrix is diagonal and we recover the well known result (see e.g. \cite{Gromov:2016itr,Kazama:2013rya,Jiang:2015lda} for recent derivations)
\beq
\label{mxsu2}
    M_\svx = \frac{1}{\langle\svx\ket{\svx}}\propto \prod_{\alpha=1}^L(\svx_\alpha-\theta_\alpha)\frac{\prod_{\beta<\alpha}(2(\theta_\beta-\theta_\alpha)-(\svx_\beta-\svx_\alpha))}{\prod_{\beta<\alpha}(\theta_\beta-\theta_\alpha)}
\eeq
where we omit the proportionality factor which is independent of $\svx$'s or $\theta$'s. Importantly, by carefully analysing the expansion of the FSoV determinant for $SU(3)$ and in fact for any $SU(N)$ one can find an explicit formula for the measure $M_{\svy,\svx}$ as well. It no longer gives a diagonal matrix and has a roughly similar but noticeably more involved combinatorial structure compared to $SU(2)$. We refer the reader to \cite{Gromov:2020fwh} for its explicit form. 

In summary, the key result is that FSoV gives determinant representations for wavefunction overlaps, as well as providing an explicit form of the SoV measure for any $SU(N)$ -- resolving an important and longstanding open problem.


\subsection{Twist-independent SoV basis from companion frame}
\label{sec:comp}

As mentioned above, the SoV measure $M_{\svy,\svx}$ turns out not to depend on the twist of the spin chain\footnote{ or the dependence is only through an overall trivial factor, identical for all matrix elements $M_{\svy,\svx}$}. As the measure $M_{\svy,\svx}$ is the inverse matrix to the matrix of overlaps $\bra{\svx}\svy\rangle$, this suggests that there is a way to choose the SoV bases $\bra{\svx}$ and $\ket{\svy}$ in such a way that they do not depend on the twist. Indeed this is possible to do as we will now explain. 

The main idea is to bring the twist matrix, originally chosen to be diagonal $g={\rm diag}(\lambda_1,\dots,\lambda_N)$, to the form known as companion twist matrix, denoted by $\Lambda$. It is defined by
\beq
    \Lambda_{ij}=(-1)^{j-1}\chi_j\delta_{i,1}+\delta_{i,j+1}
\eeq
where the symmetric polynomials $\chi_j$ are defined by
\beq\label{chid}
    \prod_{k=1}^N(t+\lambda_k)=\sum_{j=0}^N\chi_jt^{N-j}
\eeq
For example, for $SU(2)$ it reads
\beq
    \Lambda=\begin{pmatrix}
        \lambda_1+1/\lambda_1 & -1
        \\ 1 & 0
    \end{pmatrix}
\eeq
and for $SU(3)$ (with $\lambda_3=1/(\lambda_1\lambda_2)$)
\beq
        \Lambda=\begin{pmatrix}
        \lambda_1+\lambda_2+\lambda_3 & -\frac{1}{\lambda_1}-\frac{1}{\lambda_2}-\frac{1}{\lambda_3} & 1 \\ 1 & 0 & 0
        \\ 0&1 & 0
    \end{pmatrix}
\eeq
In fact $\Lambda$ has the same eigenvalues as the original diagonal twist $g$, and the transformation to the companion form can be achieved by a similarity transformation in the auxiliary space:
\beq
    g=S\Lambda S^{-1}
\eeq
The matrix $S$ here can be chosen as (see \cite{Gromov:2020fwh,Gromov:2022waj})
\beq
\label{smat}
    (S^{-1})_{ij}=\lambda_j^{N-i}
\eeq
Using the $GL(N)$ invariance of the R-matrix, one can in fact transfer this transformation onto the physical Hilbert space instead. Concretely, let us define a new monodromy matrix with the companion twist $\Lambda$ used instead of $g$ in \eq{mond},
\beq
    T^{\rm comp}(u)=T(u)g^{-1}\Lambda
\eeq
One can show that the corresponding transfer matrices are related by a global similarity transformation on the Hilbert space,
\beq
    \tr T^{\rm comp}=\Pi \; \tr T \; \Pi^{-1}
\eeq
where $\Pi=S_1,\dots S_L$ is the transformation $S$ implemented on each site of the chain. As a result, e.g. the eigenstates of the two transfer matrices are related by the same transformation,
 \beq
     \ket{\Psi^{\rm comp}}=\Pi\ket{\Psi^{\rm diag}}
 \eeq
Accordingly, if the eigenstates $\ket{\Psi^{\rm comp}}$ factorise in some SoV basis, then applying the same global rotation to this basis will give the one for the original chain. As overlaps of ket and bra states are always invariant under such a transformation, the SoV measure can be equivalently computed for either of the two cases. 

Now, the whole point of this construction is the remarkable fact that the SoV bases $\ket{\svy^{\rm comp}}$ and $\bra{\svx^{\rm comp}}$ that factorise the eigenstates of $T^{\rm comp}$ are independent of the twist. This follows from the observation that the $B$- and $C$-operators computed with this twist turn out not to depend on the twist eigenvalues as established in \cite{Ryan:2018fyo} (and extended in \cite{Gromov:2020fwh}). For example, the B-operator for $SU(2)$ in the companion frame is simply the untwisted A-operator, diagonalised in the well known Gelfand-Tsetlin basis which also plays a key role for higher rank cases (see \cite{Ryan:2018fyo} for more on this). As a result, the overlaps of these SoV states and likewise the SoV measure will be twist-independent. 

The fact that the SoV bases can be chosen in a way that does not depend on the twist plays a remarkable role in the key application of SoV -- computation of correlators. We review this in the next section.



\section{Correlators from SoV}

In this section we describe the application of the SoV formalism discussed above to computation of correlators in higher rank spin chains. We will see how it leads to new determinant formulas for a wide range of observables.

\label{sec:corr}

\subsection{Example: form factors of derivatives}

Let us start (following \cite{Cavaglia:2018lxi, Cavaglia:2019pow}) with a relatively simple but important example of a rather special type of observable effectively computable via SoV -- namely, diagonal form factors of the type ${\bra{\Psi}\frac{\d \hat I}{\d p}|\Psi\rangle}$ where $\hat I$ are some integrals of motion and $p$ is a parameter such as a twist or inhomogeneity. Using a simple argument similar to the Hellmann-Feynman theorem in quantum mechanics, one can express them in terms of derivatives of the corresponding eigenvalues, i.e.
\beq
\label{ffac}
    \frac{\bra{\Psi}\frac{\d \hat I_{b,\beta-1}}{\d p}|\Psi\rangle}{\bra{\Psi}\Psi\rangle}=\frac{\d I_{b,\beta-1}}{\d p}
\eeq
Notice, however, that the derivatives of integrals of motion are \textit{not} diagonalised by the eigenstates $\ket{\Psi}$. Consequently, the lhs of \eq{ffac} is a nontrivial matrix element. Below we will see how to express it in terms of Q-functions in the SoV approach. The usefulness of this result is twofold. First, it is already nontrivial that one can write this quantity in terms of objects evaluated at a single fixed value of the parameter $p$ (also without derivatives or infinitesimal differences etc). Second, the result reveals important structures which serve as the starting point for wider generalisations.

The argument is rather similar to usual quantum mechanical perturbation theory.  As usual, let us focus on the $SU(3)$ case. We start by considering a small variation of the parameter, under which Q-functions change as $Q\to Q+\delta Q$ and the coefficients in the Baxter operator \eq{od3} will change as well, so $\hat O^\dagger \to \hat O^\dagger+\delta \hat O^\dagger$. Then we have, to first order in the variations,
\beq
    0=\langle Q_1(\hat O^\dagger+\delta\hat O^\dagger)(Q_{1,a+1}+\delta Q_{1,a+1})\rangle_\alpha=\langle Q_1\hat O^\dagger\delta Q_{1,a+1}\rangle_\alpha+\langle Q_1\delta\hat O^\dagger Q_{1,a+1}\rangle_\alpha \ .
\eeq
Using the key adjointness property \eq{oad}, we see that the first bracket here is identically zero, so we conclude that
\beq
\label{ovar1}
    \langle Q_1\delta\hat O^\dagger Q_{1,a+1}\rangle_\alpha =0 \ .
\eeq
Let us for simplicity consider the variation in one of the twist parameters (variations in e.g. inhomogeneities can be derived in a similar way). Then, explicitly we find that the variation of $\hat O^\dagger$ has the form
\beq
\d_p \hat { O }^\dagger = \sum_{(b,\beta)}(-1)^{b+1}\d_p I_{b,\beta-1} u^{\beta-1}{D}^{-2b+3}
-\hat Y_p
\eeq
where
\beq
\label{Yvar}
\hat Y_p=
-\sum_{b}(-1)^{b+1}\d_p I_{b,L} u^{L}{D}^{-2b+3} \;.
\eeq
We denoted by $I_{b,L}$ the leading coefficient in the transfer matrices, $I_{b,L}=\chi_b(\lambda_1,\lambda_2,\lambda_3)$. 
Next we plug this into \eq{ovar1} which gives a linear system for the variations $\d_p I_{b,\beta-1}$:
\beq\la{ABN1}
\sum_{(b,\beta)}m_{(a,\alpha),(b,\beta)}(-1)^{b+1}\d_p I_{b,\beta-1} =
y_{(a,\alpha)}\;\;,\;\;y_{(a,\alpha)}\equiv \langle Q_1\;\hat Y_p\circ Q_{1,a+1} \rangle_\alpha\;
\eeq
Here $m_{(a,\alpha),(b,\beta)}$ is the same matrix appearing in the scalar product we derived using functional SoV above in \eq{detsl3}, with the two states $A$ and $B$ taken to be the same,
\beq
m_{(a,\alpha),(b,\beta)}=\det_{(a,\alpha),(b,\beta)}\langle Q_1^A \;u^{\beta-1}\; {\cal D}^{3-2b} Q_{1,a+1}^B \rangle_\alpha
\eeq
We can write its solution using the Cramer's rule as
\beq
\label{Ivar}
\d_p I_{b',\beta'-1}=(-1)^{b'+1}\frac{\det_{(a,\alpha),(b,\beta)}\tilde m_{(a,\alpha),(b,\beta)}}{\det_{(a,\alpha),(b,\beta)}m_{(a,\alpha),(b,\beta)}} \;,
\eeq
Here $\tilde m_{(a,\alpha),(b,\beta)}$ is the matrix
$m_{(a,\alpha),(b,\beta)}$ with the column $(b',\beta')$ replaced with $y_{(a,\alpha)}$ given in \eq{ABN1}. Nicely, this gives a determinant representation for the variation of the integrals of motion and hence also for  the form factor \eq{ffac}. 


\subsection{General matrix elements of a complete set of operators}

The form factors computed above are the basic example of observables accessible via SoV. Using a variety of tricks, it has been possible to obtain similar determinant formulas for a much wider range of correlators -- see \cite{Gromov:2020fwh} for various examples, including e.g. overlaps for states with different twists (that have been much studied via other approaches as well, see  the reviews \cite{Slavnov:2018kfx}, \cite{Slavnov:2019hdn}) and on-shell or off-shell form factors of $B$ and $C$ operators. One may also, for example, define 'off-shell' states by acting with the B-operator on the vacuum as in \eq{psu3} where the parameters $u_k$ are kept arbitrary instead of being the Bethe roots -- then overlaps of such states (on-shell or off-shell) also have a compact determinant form. 

Eventually these results led to an even more powerful observation in \cite{Gromov:2022waj} that there is a \textit{complete} set of operators all of whose matrix elements (including non-diagonal ones!) are given by compact SoV expressions. This is one of the main outcomes to date of the SoV program for higher rank spin chains. For the sake of readability of this pedagogical review, we will not discuss these results in full technicality here, but rather outline some key features and refer to \cite{Gromov:2022waj} for a complete presentation.

The main idea is to consider, similarly to \eq{ovar1} above, the equation 
    \beq
    \langle Q_1\hat O^\dagger Q_{1,a+1}\rangle_\alpha =0 \ .
\eeq
which is trivially satisfied due to the Baxter equation $O^\dagger Q_{1,a+1}=0$. This means that one can express the integrals of motion themselves as solutions to a linear system with the same matrix  as we had in \eq{su3lin} before. Next comes the most important part -- using the fact that the SoV basis is independent of the twist, as well as special features of the companion twist frame, one can 'upgrade' this result in several ways. Eventually one can find  all matrix elements for a complete set of operators that  were called principal operators in \cite{Gromov:2022waj}.


Roughly speaking, the principal operators are certain entries of $T$, as well as their particular quadratic, cubic etc combinations. As an example, let us give here explicitly a few of the principal operators for $SU(N)$. They have the simplest form in the companion twist frame discussed in section \ref{sec:comp}, whereas for usual diagonal twist they can be written conveniently by introducing notation similar to $T^{\g}$ from section \ref{sec:tg3}. Namely, we use the matrix $S$ from \eq{smat} to do a similarity transformation in the auxiliary space, defining the new monodromy matrix by (see section 7.2 of \cite{Gromov:2022waj})
\beq
    T^{S}(u)=S^{-1}T(u)S
\eeq
Then examples of principal operators are:
\beq
    P_{1,i}(u)=(-1)^{N-i}\frac{T^S_{iN}(u)}{\chi_N} \ , \ \ i=1,2,\dots,N
\eeq
\beq
    P_{1,0}(u)=\sum_{i=1}^{N-1}\(T^S_{ii}(u)-(-1)^{N-i}\frac{\chi_i}{\chi_N}T^S_{iN}(u)\)
\eeq
where the coefficients $\chi_k$ are defined in \eq{chid}. There are also many more principal operators, given by higher degree polynomials in entries of the monodromy matrix. For all of these operators, their matrix elements are given by simple SoV-type determinants. Even more complicated observables, for example some antisymmetrised correlators of multiple insertions of these operators, are also under control, and we refer to \cite{Gromov:2022waj} for details and extensions. 

In general, the calculation of correlators is one of the main expected applications of SoV. We expect that the realisation of SoV as well as FSoV methods and knowledge of the SoV measure, finally available now for higher rank spin chains, should open the way to efficient exploration of a wide  range of important observables in the future.


\section{Conclusions and future directions}
\label{sec:concl}

In this review we described how SoV can be realised for higher rank $SU(N)$ spin chains, and showcased key applications such as a new compact construction of eigenstates and new determinant representations for many correlators and overlaps. With these advances, the way is open to attack a wide range of questions. Let us list here some of them.

\begin{itemize}

\item It would be illuminating to better understand the representation theory meaning of various components of the new SoV constructions (e.g. the SoV measure and the self-adjointness of Baxter operators). While so far the approach has been partially `experimental', the simplicity of final results suggests a deeper algebraic interpretation. Representation-theoretic objects such as Gelfand-Tsetlin bases already feature prominently in SoV, and it would be interesting to uncover more  underlying algebraic structures. Let us note that a more invariant description of the SoV B-operator was already discussed in \cite{Chervov:2007bb, smirnov2002separation}.

\item A key future direction will be to explore in more detail the full implications of compact determinant representations that SoV gives for correlators and other quantities in higher rank models, as well as extending them to more observables and comparing with the results obtained by usual Bethe ansatz. 

\item It is also important to study the thermodynamic limit of various correlators in higher-rank models, as was discussed  for rank 1 in e.g. \cite{borot2016asymptotic,borot2015large} where the problem is related to matrix model-like integrals and $\beta$-ensembles. The higher rank case appears to give novel kinds of integrals of this type that call for active exploration. It is also expected that the available explicit form of the SoV measure will help to progress here, like in the $SU(2)$ case \cite{Niccoli:2020zla}.

\item A highly interesting related direction is applying the powerful SoV tools to computation of non-local observables such as entanglement entropy \cite{keating2004random,Castro-Alvaredo:2010scy}, exploration of quantum quenches \cite{refael2004entanglement} and integrable hydrodynamics \cite{Castro-Alvaredo:2016cdj,DeNardis:2018omc}.

\item A highly promising application of SoV is for N=4 super Yang-Mills (SYM) theory \cite{Beisert:2010jr} where it has already shown to provide massive simplifications for certain correlators in SYM \cite{Cavaglia:2018lxi, Giombi:2018qox, Bercini:2022jxo}\footnote{See also \cite{McGovern:2019sdd,Caetano:2020dyp} for related observations}. The crucial insight is that the Q-functions, into which the states are expected to factorise, are under excellent control for all values of the coupling from the Quantum Spectral Curve equations \cite{Gromov:2013pga} that determine the model's spectrum. Establishing this framework rigorously and extending it to general correlators is a key future direction, opening the way to the complete non-perturbative solution of the SYM theory.

\item A closely linked direction is exploring similar questions in remarkable simplified version of SYM known as fishnet CFTs \cite{Gurdogan:2015csr}. Here calculation of key observables maps to spin chains based on the conformal group, in a principal series representation\footnote{see \cite{Derkachov:2018ewi} for recent advances in using the SoV B-operator in a similar setup for higher rank cases} where Bethe ansatz is not applicable but SoV should reveal its power. A large part of the SoV groundwork for this case was recently established in \cite{Cavaglia:2021mft} for 4d fishnets, already leading to novel compact results for correlators. Pursuing this program will open the way to computing yet more involved and rich observables in fishnet and other theories, and will be crucial for uncovering similar structures for SYM as well.

\item A related set of questions for fishnet theories was explored in \cite{Derkachov:2018rot,Derkachov:2019tzo,Derkachov:2021rrf,Derkachov:2021ufp} which use methods in many ways similar to the SoV we discuss here in order to attack nontrivial field theory calculations, leading to numerous important results. This approach is somewhat different from what we reviewed in this article, as roughly speaking factorisation is achieved not for the eigenstates of the usual transfer matrix but rather for eigenstates of its principal minors (which is often easier). Besides providing important new field theory observables, these results should also help to guide the usual SoV realisation for the standard transfer matrix eigenstates in these conformal spin chains. 

\item Spin chains or Gaudin models based on the conformal $SO(D,2)$ group symmetry have been recently shown to provide interesting field theory quantities such as multiloop Feynman integrals \cite{Chicherin:2017cns, Chicherin:2017frs, Loebbert:2022nfu, Kazakov:2023nyu, Levkovich-Maslyuk:2024zdy} and higher-point conformal blocks \cite{Buric:2020dyz,Buric:2021ywo} as their eigenstates. These models are typically not treatable via Bethe ansatz, so it would be especially interesting to utilise higher-rank SoV methods to seek new representations of these observables.

\item It would be interesting to further explore the role of SoV in integrable systems realised in 4d $N=2$ (and related) gauge theories within the approach of \cite{Nekrasov:2009rc} and its extensions. This setting can also provide new insights on SoV itself, see e.g. \cite{Jeong:2024onv} for recent results for higher-rank Gaudin models. A related question is clarifying the role of SoV in relation to bispectral dualities between integrable systems. 

\item A natural question is looking for extensions of the new SoV methods from the rational to the trigonometric and elliptic cases (see \cite{Maillet:2018rto} for initial results), as well as to boundary problems in spin chains \cite{Arnaudon:2004sd} (see \cite{Ekhammar:2023iph} for recent progress). For the supersymmetric case also only partial results are known so far despite its importance in statistical physics (e.g. for the super $t-J$ model) and in super Yang-Mills theory. Other important extensions include various Gaudin models and long-range spin chains like the Calogero-Sutherland model.

\item We have seen here that determinant representations for overlaps in SoV follow essentially from solving a linear system of equations. A intriguing related observation was recently made \cite{Belliard:2019bfz} in the context of more traditional Bethe ansatz methods, and it would be interesting to connect the two.

\end{itemize}

\section*{Acknowledgements}
I am grateful to  A.~Cavaglia, N.~Gromov, P.~Ryan, G.~Sizov and D.~Volin for enjoyable collaborations on various aspects of separation of variables. I also thank  S. Belliard, A. Chervov, S. Ekhammar, A. Hutsalyuk, J. Lamers, G. Lefundes, A. Liashyk, D. Serban and B. Vicedo for many related discussions. My work was supported by STFC grant APP69281.

	\appendix
	
\section{The operators B and C for any $SU(N)$}
\label{app:bc}

Here we give the explicit form of the $B$- and $C$-operators for any $SU(N)$. We follow \cite{Gromov:2016itr}, \cite{Gromov:2019wmz}, \cite{Gromov:2020fwh}.

First, the B-operator reads
\begin{equation}\label{Bdef}
    B=\sum_{j_k}T\left[^{j_1}_{N}\right]\bT^{[-2]}\left[^{j_2}_{j_1,N}\right]T^{[-4]}\left[^{j_3}_{j_2,N}\right]\dots T^{[-2N+4]}\left[^{12\dots N-1}_{j_{N-2},N}\right]
\end{equation}
Here $j_k=\{j_k^1,\dots,j_k^k\}$, $k=1,2,\dots,N-2$ is a multi-index and the sum runs over all values with $1\leq j_k^1<j_k^2<\dots <j_k^k\leq N-1$. The  $\bC$ operator in the same notation is given by
\begin{equation}
    C=\sum_{j_k}T\left[^{12\dots N-1}_{j_{N-2},N}\right]\dots T\left[^{j_3}_{j_2,N}\right]T\left[^{j_2}_{j_1,N}\right]T\left[^{j_1}_{N}\right]\;.
\end{equation}

\bibliographystyle{JHEP.bst}
 \bibliography{SoVbib}

\providecommand{\href}[2]{#2}\begingroup\raggedright\begin{thebibliography}{100}

\bibitem{Beisert:2010jr}
N.~Beisert et~al., \emph{{Review of AdS/CFT Integrability: An Overview}}, \href{https://doi.org/10.1007/s11005-011-0529-2}{\emph{Lett. Math. Phys.} {\bfseries 99} (2012) 3} [\href{https://arxiv.org/abs/1012.3982}{{\ttfamily 1012.3982}}].

\bibitem{Sklyanin:1984sb}
E.K.~Sklyanin, \emph{{The Quantum Toda Chain}}, {\emph{Lect. Notes Phys.} {\bfseries 226} (1985) 196}.

\bibitem{Sklyanin:1987ih}
E.K.~Sklyanin, \emph{{Separation of variables in the Gaudin model}}, \href{https://doi.org/10.1007/BF01840429}{\emph{Zap. Nauchn. Semin.} {\bfseries 164} (1987) 151}.

\bibitem{Sklyanin:1991ss}
E.K.~Sklyanin, \emph{{Quantum inverse scattering method. Selected topics}},  \href{https://arxiv.org/abs/hep-th/9211111}{{\ttfamily hep-th/9211111}}.

\bibitem{Sklyanin:1992eu}
E.K.~Sklyanin, \emph{{Separation of variables in the classical integrable SL(3) magnetic chain}}, \href{https://doi.org/10.1007/BF02096572}{\emph{Commun. Math. Phys.} {\bfseries 150} (1992) 181} [\href{https://arxiv.org/abs/hep-th/9211126}{{\ttfamily hep-th/9211126}}].

\bibitem{Sklyanin:1992sm}
E.K.~Sklyanin, \emph{{Separation of variables in the quantum integrable models related to the Yangian Y[sl(3)]}}, \href{https://doi.org/10.1007/BF02362784}{\emph{Zap. Nauchn. Semin.} {\bfseries 205} (1993) 166} [\href{https://arxiv.org/abs/hep-th/9212076}{{\ttfamily hep-th/9212076}}].

\bibitem{Sklyanin:1995bm}
E.K.~Sklyanin, \emph{{Separation of variables - new trends}}, \href{https://doi.org/10.1143/PTPS.118.35}{\emph{Prog. Theor. Phys. Suppl.} {\bfseries 118} (1995) 35} [\href{https://arxiv.org/abs/solv-int/9504001}{{\ttfamily solv-int/9504001}}].

\bibitem{Niccoli:2011nj}
G.~Niccoli, \emph{{Completeness of Bethe Ansatz by Sklyanin SOV for Cyclic Representations of Integrable Quantum Models}}, \href{https://doi.org/10.1007/JHEP03(2011)123}{\emph{JHEP} {\bfseries 03} (2011) 123} [\href{https://arxiv.org/abs/1102.1694}{{\ttfamily 1102.1694}}].

\bibitem{Niccoli:2012ci}
G.~Niccoli, \emph{{Antiperiodic spin-1/2 XXZ quantum chains by separation of variables: Complete spectrum and form factors}}, \href{https://doi.org/10.1016/j.nuclphysb.2013.01.017}{\emph{Nucl. Phys. B} {\bfseries 870} (2013) 397} [\href{https://arxiv.org/abs/1205.4537}{{\ttfamily 1205.4537}}].

\bibitem{Grosjean:2012ay}
N.~Grosjean and G.~Niccoli, \emph{{The $\tau_2$-model and the chiral Potts model revisited: completeness of Bethe equations from Sklyanin's SOV method}}, \href{https://doi.org/10.1088/1742-5468/2012/11/P11005}{\emph{J. Stat. Mech.} {\bfseries 1211} (2012) P11005} [\href{https://arxiv.org/abs/1205.4614}{{\ttfamily 1205.4614}}].

\bibitem{Niccoli:2012vq}
G.~Niccoli, \emph{{Form factors and complete spectrum of XXX antiperiodic higher spin chains by quantum separation of variables}}, \href{https://doi.org/10.1063/1.4807078}{\emph{J. Math. Phys.} {\bfseries 54} (2013) 053516} [\href{https://arxiv.org/abs/1206.2418}{{\ttfamily 1206.2418}}].

\bibitem{Niccoli:2012pi}
G.~Niccoli, \emph{{Non-diagonal open spin-1/2 XXZ quantum chains by separation of variables: Complete spectrum and matrix elements of some quasi-local operators}}, \href{https://doi.org/10.1088/1742-5468/2012/10/P10025}{\emph{J. Stat. Mech.} {\bfseries 1210} (2012) P10025} [\href{https://arxiv.org/abs/1206.0646}{{\ttfamily 1206.0646}}].

\bibitem{Niccoli:2012mq}
G.~Niccoli, \emph{{Antiperiodic dynamical 6-vertex and periodic 8-vertex models I: Complete spectrum by SOV and matrix elements of the identity on separate states}}, \href{https://doi.org/10.1088/1751-8113/46/7/075003}{\emph{J. Phys. A} {\bfseries 46} (2013) 075003} [\href{https://arxiv.org/abs/1207.1928}{{\ttfamily 1207.1928}}].

\bibitem{Faldella:2013qha}
S.~Faldella, N.~Kitanine and G.~Niccoli, \emph{{The complete spectrum and scalar products for the open spin-1/2 XXZ quantum chains with non-diagonal boundary terms}}, \href{https://doi.org/10.1088/1742-5468/2014/01/P01011}{\emph{J. Stat. Mech.} {\bfseries 1401} (2014) P01011} [\href{https://arxiv.org/abs/1307.3960}{{\ttfamily 1307.3960}}].

\bibitem{Faldella:2013qca}
S.~Faldella and G.~Niccoli, \emph{{SOV approach for integrable quantum models associated with general representations on spin-1/2 chains of the 8-vertex reflection algebra}}, \href{https://doi.org/10.1088/1751-8113/47/11/115202}{\emph{J. Phys. A} {\bfseries 47} (2014) 115202} [\href{https://arxiv.org/abs/1307.5531}{{\ttfamily 1307.5531}}].

\bibitem{Kitanine:2014swa}
N.~Kitanine, J.M.~Maillet and G.~Niccoli, \emph{{Open spin chains with generic integrable boundaries: Baxter equation and Bethe ansatz completeness from separation of variables}}, \href{https://doi.org/10.1088/1742-5468/2014/05/P05015}{\emph{J. Stat. Mech.} {\bfseries 1405} (2014) P05015} [\href{https://arxiv.org/abs/1401.4901}{{\ttfamily 1401.4901}}].

\bibitem{Niccoli:2014sfa}
G.~Niccoli and V.~Terras, \emph{{Antiperiodic XXZ chains with arbitrary spins: Complete eigenstate construction by functional equations in separation of variables}}, \href{https://doi.org/10.1007/s11005-015-0759-9}{\emph{Lett. Math. Phys.} {\bfseries 105} (2015) 989} [\href{https://arxiv.org/abs/1411.6488}{{\ttfamily 1411.6488}}].

\bibitem{Kitanine:2015jna}
N.~Kitanine, J.M.~Maillet, G.~Niccoli and V.~Terras, \emph{{On determinant representations of scalar products and form factors in the SoV approach: the XXX case}}, \href{https://doi.org/10.1088/1751-8113/49/10/104002}{\emph{J. Phys. A} {\bfseries 49} (2016) 104002} [\href{https://arxiv.org/abs/1506.02630}{{\ttfamily 1506.02630}}].

\bibitem{Levy-Bencheton:2015mia}
D.~Levy-Bencheton, G.~Niccoli and V.~Terras, \emph{{Antiperiodic dynamical 6-vertex model by separation of variables II: Functional equations and form factors}}, \href{https://doi.org/10.1088/1742-5468/2016/03/033110}{\emph{J. Stat. Mech.} {\bfseries 1603} (2016) 033110} [\href{https://arxiv.org/abs/1507.03404}{{\ttfamily 1507.03404}}].

\bibitem{Kitanine:2016pvg}
N.~Kitanine, J.M.~Maillet, G.~Niccoli and V.~Terras, \emph{{The open XXX spin chain in the SoV framework: scalar product of separate states}}, \href{https://doi.org/10.1088/1751-8121/aa6cc9}{\emph{J. Phys. A} {\bfseries 50} (2017) 224001} [\href{https://arxiv.org/abs/1606.06917}{{\ttfamily 1606.06917}}].

\bibitem{Kitanine:2018gki}
N.~Kitanine, J.M.~Maillet, G.~Niccoli and V.~Terras, \emph{{The open XXZ spin chain in the SoV framework: scalar product of separate states}}, \href{https://doi.org/10.1088/1751-8121/aae76f}{\emph{J. Phys. A} {\bfseries 51} (2018) 485201} [\href{https://arxiv.org/abs/1807.05197}{{\ttfamily 1807.05197}}].

\bibitem{Niccoli:2020zla}
G.~Niccoli, H.~Pei and V.~Terras, \emph{{Correlation functions by separation of variables: the XXX spin chain}}, \href{https://doi.org/10.21468/SciPostPhys.10.1.006}{\emph{SciPost Phys.} {\bfseries 10} (2021) 006} [\href{https://arxiv.org/abs/2005.01334}{{\ttfamily 2005.01334}}].

\bibitem{Niccoli:2024usi}
G.~Niccoli and V.~Terras, \emph{{The open XYZ spin 1/2 chain: Separation of Variables and scalar products for boundary fields related by a constraint}},  \href{https://arxiv.org/abs/2402.04112}{{\ttfamily 2402.04112}}.

\bibitem{Derkachov:2001yn}
S.E.~Derkachov, G.P.~Korchemsky and A.N.~Manashov, \emph{{Noncompact Heisenberg spin magnets from high-energy QCD: 1. Baxter Q operator and separation of variables}}, \href{https://doi.org/10.1016/S0550-3213(01)00457-6}{\emph{Nucl. Phys. B} {\bfseries 617} (2001) 375} [\href{https://arxiv.org/abs/hep-th/0107193}{{\ttfamily hep-th/0107193}}].

\bibitem{Derkachov:2002wz}
S.E.~Derkachov, G.P.~Korchemsky, J.~Kotanski and A.N.~Manashov, \emph{{Noncompact Heisenberg spin magnets from high-energy QCD. 2. Quantization conditions and energy spectrum}}, \href{https://doi.org/10.1016/S0550-3213(02)00842-8}{\emph{Nucl. Phys. B} {\bfseries 645} (2002) 237} [\href{https://arxiv.org/abs/hep-th/0204124}{{\ttfamily hep-th/0204124}}].

\bibitem{Derkachov:2002pb}
S.E.~Derkachov, G.P.~Korchemsky and A.N.~Manashov, \emph{{Noncompact Heisenberg spin magnets from high-energy QCD. 3. Quasiclassical approach}}, \href{https://doi.org/10.1016/S0550-3213(03)00340-7}{\emph{Nucl. Phys. B} {\bfseries 661} (2003) 533} [\href{https://arxiv.org/abs/hep-th/0212169}{{\ttfamily hep-th/0212169}}].

\bibitem{Derkachov:2002tf}
S.E.~Derkachov, G.P.~Korchemsky and A.N.~Manashov, \emph{{Separation of variables for the quantum SL(2,R) spin chain}}, \href{https://doi.org/10.1088/1126-6708/2003/07/047}{\emph{JHEP} {\bfseries 07} (2003) 047} [\href{https://arxiv.org/abs/hep-th/0210216}{{\ttfamily hep-th/0210216}}].

\bibitem{Derkachov:2003qb}
S.E.~Derkachov, G.P.~Korchemsky and A.N.~Manashov, \emph{{Baxter Q operator and separation of variables for the open SL(2,R) spin chain}}, \href{https://doi.org/10.1088/1126-6708/2003/10/053}{\emph{JHEP} {\bfseries 10} (2003) 053} [\href{https://arxiv.org/abs/hep-th/0309144}{{\ttfamily hep-th/0309144}}].

\bibitem{Guica:2017mtd}
M.~Guica, F.~Levkovich-Maslyuk and K.~Zarembo, \emph{{Integrability in dipole-deformed $\boldsymbol{\mathcal{N}=4}$ super Yang\textendash{}Mills}}, \href{https://doi.org/10.1088/1751-8121/aa8491}{\emph{J. Phys. A} {\bfseries 50} (2017) 39} [\href{https://arxiv.org/abs/1706.07957}{{\ttfamily 1706.07957}}].

\bibitem{Smirnov:1998kv}
F.A.~Smirnov, \emph{{Quasiclassical study of form-factors in finite volume}},  \href{https://arxiv.org/abs/hep-th/9802132}{{\ttfamily hep-th/9802132}}.

\bibitem{Negro:2013wga}
S.~Negro and F.~Smirnov, \emph{{On one-point functions for sinh-Gordon model at finite temperature}}, \href{https://doi.org/10.1016/j.nuclphysb.2013.06.023}{\emph{Nucl. Phys. B} {\bfseries 875} (2013) 166} [\href{https://arxiv.org/abs/1306.1476}{{\ttfamily 1306.1476}}].

\bibitem{Scott:1994dz}
D.R.D.~Scott, \emph{{Classical functional Bethe ansatz for SL(N): Separation of variables for the magnetic chain}}, \href{https://doi.org/10.1063/1.530712}{\emph{J. Math. Phys.} {\bfseries 35} (1994) 5831} [\href{https://arxiv.org/abs/hep-th/9403030}{{\ttfamily hep-th/9403030}}].

\bibitem{gekhtman1995separation}
M.~Gekhtman, \emph{Separation of variables in the classical $\backslash$rmsl(n) magnetic chain}, .

\bibitem{smirnov2002separation}
F.A.~Smirnov, \emph{Separation of variables for quantum integrable models related to},  in \emph{MathPhys Odyssey 2001: Integrable Models and Beyond In Honor of Barry M. McCoy}, pp.~455--465, Springer (2002).

\bibitem{Gromov:2016itr}
N.~Gromov, F.~Levkovich-Maslyuk and G.~Sizov, \emph{{New Construction of Eigenstates and Separation of Variables for SU(N) Quantum Spin Chains}}, \href{https://doi.org/10.1007/JHEP09(2017)111}{\emph{JHEP} {\bfseries 09} (2017) 111} [\href{https://arxiv.org/abs/1610.08032}{{\ttfamily 1610.08032}}].

\bibitem{Liashyk:2018qfc}
A.~Liashyk and N.A.~Slavnov, \emph{{On Bethe vectors in $\mathfrak{gl}_3$-invariant integrable models}}, \href{https://doi.org/10.1007/JHEP06(2018)018}{\emph{JHEP} {\bfseries 06} (2018) 018} [\href{https://arxiv.org/abs/1803.07628}{{\ttfamily 1803.07628}}].

\bibitem{Gromov:2018cvh}
N.~Gromov and F.~Levkovich-Maslyuk, \emph{{New Compact Construction of Eigenstates for Supersymmetric Spin Chains}}, \href{https://doi.org/10.1007/JHEP09(2018)085}{\emph{JHEP} {\bfseries 09} (2018) 085} [\href{https://arxiv.org/abs/1805.03927}{{\ttfamily 1805.03927}}].

\bibitem{Chervov:2007bb}
A.~Chervov and G.~Falqui, \emph{{Manin matrices and Talalaev's formula}}, \href{https://doi.org/10.1088/1751-8113/41/19/194006}{\emph{J. Phys. A} {\bfseries 41} (2008) 194006} [\href{https://arxiv.org/abs/0711.2236}{{\ttfamily 0711.2236}}].

\bibitem{Maillet:2018bim}
J.M.~Maillet and G.~Niccoli, \emph{{On quantum separation of variables}}, \href{https://doi.org/10.1063/1.5050989}{\emph{J. Math. Phys.} {\bfseries 59} (2018) 091417} [\href{https://arxiv.org/abs/1807.11572}{{\ttfamily 1807.11572}}].

\bibitem{Maillet:2018czd}
J.M.~Maillet and G.~Niccoli, \emph{{Complete spectrum of quantum integrable lattice models associated to Y(gl(n)) by separation of variables}}, \href{https://doi.org/10.21468/SciPostPhys.6.6.071}{\emph{SciPost Phys.} {\bfseries 6} (2019) 071} [\href{https://arxiv.org/abs/1810.11885}{{\ttfamily 1810.11885}}].

\bibitem{Ryan:2018fyo}
P.~Ryan and D.~Volin, \emph{{Separated variables and wave functions for rational gl(N) spin chains in the companion twist frame}}, \href{https://doi.org/10.1063/1.5085387}{\emph{J. Math. Phys.} {\bfseries 60} (2019) 032701} [\href{https://arxiv.org/abs/1810.10996}{{\ttfamily 1810.10996}}].

\bibitem{Maillet:2018rto}
J.M.~Maillet and G.~Niccoli, \emph{{Complete spectrum of quantum integrable lattice models associated to $\mathcal{U}_{q} (\widehat{gl_{n}})$ by separation of variables}}, \href{https://doi.org/10.1088/1751-8121/ab2930}{\emph{J. Phys. A} {\bfseries 52} (2019) 315203} [\href{https://arxiv.org/abs/1811.08405}{{\ttfamily 1811.08405}}].

\bibitem{Maillet:2019nsy}
J.M.~Maillet and G.~Niccoli, \emph{{On quantum separation of variables beyond fundamental representations}}, \href{https://doi.org/10.21468/SciPostPhys.10.2.026}{\emph{SciPost Phys.} {\bfseries 10} (2021) 026} [\href{https://arxiv.org/abs/1903.06618}{{\ttfamily 1903.06618}}].

\bibitem{Maillet:2019hdq}
J.M.~Maillet and G.~Niccoli, \emph{{On Separation of Variables for Reflection Algebras}}, \href{https://doi.org/10.1088/1742-5468/ab357a}{\emph{J. Stat. Mech.} {\bfseries 1909} (2019) 094020} [\href{https://arxiv.org/abs/1904.00852}{{\ttfamily 1904.00852}}].

\bibitem{Maillet:2019ayx}
J.M.~Maillet, G.~Niccoli and L.~Vignoli, \emph{{Separation of variables bases for integrable $gl_{\mathcal{M}|\mathcal{N}}$ and Hubbard models}}, \href{https://doi.org/10.21468/SciPostPhys.9.4.060}{\emph{SciPost Phys.} {\bfseries 9} (2020) 060} [\href{https://arxiv.org/abs/1907.08124}{{\ttfamily 1907.08124}}].

\bibitem{Ryan:2020rfk}
P.~Ryan and D.~Volin, \emph{{Separation of Variables for Rational $\mathfrak {gl}(\mathsf {n})$ Spin Chains in Any Compact Representation, via Fusion, Embedding Morphism and B\"acklund Flow}}, \href{https://doi.org/10.1007/s00220-021-03990-7}{\emph{Commun. Math. Phys.} {\bfseries 383} (2021) 311} [\href{https://arxiv.org/abs/2002.12341}{{\ttfamily 2002.12341}}].

\bibitem{Cavaglia:2019pow}
A.~Cavagli\`a, N.~Gromov and F.~Levkovich-Maslyuk, \emph{{Separation of variables and scalar products at any rank}}, \href{https://doi.org/10.1007/JHEP09(2019)052}{\emph{JHEP} {\bfseries 09} (2019) 052} [\href{https://arxiv.org/abs/1907.03788}{{\ttfamily 1907.03788}}].

\bibitem{Gromov:2019wmz}
N.~Gromov, F.~Levkovich-Maslyuk, P.~Ryan and D.~Volin, \emph{{Dual Separated Variables and Scalar Products}}, \href{https://doi.org/10.1016/j.physletb.2020.135494}{\emph{Phys. Lett. B} {\bfseries 806} (2020) 135494} [\href{https://arxiv.org/abs/1910.13442}{{\ttfamily 1910.13442}}].

\bibitem{smirnov2002affine}
F.~Smirnov and V.~Zeitlin, \emph{Affine jacobians of spectral curves and integrable models}, {\emph{arXiv preprint math-ph/0203037} (2002) }.

\bibitem{smirnov2002quantization}
F.A.~Smirnov and V.~Zeitlin, \emph{On the quantization of affine jacobi varieties of spectral curves},  in \emph{Statistical Field Theories}, pp.~79--89, Springer (2002).

\bibitem{Martin:2015eea}
D.~Martin and F.~Smirnov, \emph{{Problems with using separated variables for computing expectation values for higher ranks}}, \href{https://doi.org/10.1007/s11005-016-0823-0}{\emph{Lett. Math. Phys.} {\bfseries 106} (2016) 469} [\href{https://arxiv.org/abs/1506.08042}{{\ttfamily 1506.08042}}].

\bibitem{Maillet:2020ykb}
J.M.~Maillet, G.~Niccoli and L.~Vignoli, \emph{{On Scalar Products in Higher Rank Quantum Separation of Variables}}, \href{https://doi.org/10.21468/SciPostPhys.9.6.086}{\emph{SciPost Phys.} {\bfseries 9} (2020) 086} [\href{https://arxiv.org/abs/2003.04281}{{\ttfamily 2003.04281}}].

\bibitem{Gromov:2020fwh}
N.~Gromov, F.~Levkovich-Maslyuk and P.~Ryan, \emph{{Determinant form of correlators in high rank integrable spin chains via separation of variables}}, \href{https://doi.org/10.1007/JHEP05(2021)169}{\emph{JHEP} {\bfseries 05} (2021) 169} [\href{https://arxiv.org/abs/2011.08229}{{\ttfamily 2011.08229}}].

\bibitem{Gromov:2022waj}
N.~Gromov, N.~Primi and P.~Ryan, \emph{{Form-factors and complete basis of observables via separation of variables for higher rank spin chains}}, \href{https://doi.org/10.1007/JHEP11(2022)039}{\emph{JHEP} {\bfseries 11} (2022) 039} [\href{https://arxiv.org/abs/2202.01591}{{\ttfamily 2202.01591}}].

\bibitem{Ekhammar:2023iph}
S.~Ekhammar, N.~Gromov and P.~Ryan, \emph{{Boundary overlaps from Functional Separation of Variables}}, \href{https://doi.org/10.1007/JHEP05(2024)268}{\emph{JHEP} {\bfseries 05} (2024) 268} [\href{https://arxiv.org/abs/2312.11612}{{\ttfamily 2312.11612}}].

\bibitem{Gromov:2013pga}
N.~Gromov, V.~Kazakov, S.~Leurent and D.~Volin, \emph{{Quantum Spectral Curve for Planar $\mathcal{N} = 4$ Super-Yang-Mills Theory}}, \href{https://doi.org/10.1103/PhysRevLett.112.011602}{\emph{Phys. Rev. Lett.} {\bfseries 112} (2014) 011602} [\href{https://arxiv.org/abs/1305.1939}{{\ttfamily 1305.1939}}].

\bibitem{Basso:2015zoa}
B.~Basso, S.~Komatsu and P.~Vieira, \emph{{Structure Constants and Integrable Bootstrap in Planar N=4 SYM Theory}},  \href{https://arxiv.org/abs/1505.06745}{{\ttfamily 1505.06745}}.

\bibitem{Bargheer:2017nne}
T.~Bargheer, J.~Caetano, T.~Fleury, S.~Komatsu and P.~Vieira, \emph{{Handling Handles: Nonplanar Integrability in $\mathcal{N}=4$ Supersymmetric Yang-Mills Theory}}, \href{https://doi.org/10.1103/PhysRevLett.121.231602}{\emph{Phys. Rev. Lett.} {\bfseries 121} (2018) 231602} [\href{https://arxiv.org/abs/1711.05326}{{\ttfamily 1711.05326}}].

\bibitem{Vicedo:2008ryn}
B.~Vicedo, \emph{{The method of finite-gap integration in classical and semi-classical string theory}}, \href{https://doi.org/10.1088/1751-8113/44/12/124002}{\emph{J. Phys. A} {\bfseries 44} (2011) 124002} [\href{https://arxiv.org/abs/0810.3402}{{\ttfamily 0810.3402}}].

\bibitem{Kazama:2012is}
Y.~Kazama and S.~Komatsu, \emph{{Wave functions and correlation functions for GKP strings from integrability}}, \href{https://doi.org/10.1007/JHEP09(2012)022}{\emph{JHEP} {\bfseries 09} (2012) 022} [\href{https://arxiv.org/abs/1205.6060}{{\ttfamily 1205.6060}}].

\bibitem{Sobko:2013ema}
E.~Sobko, \emph{{A new representation for two- and three-point correlators of operators from sl(2) sector}}, \href{https://doi.org/10.1007/JHEP12(2014)101}{\emph{JHEP} {\bfseries 12} (2014) 101} [\href{https://arxiv.org/abs/1311.6957}{{\ttfamily 1311.6957}}].

\bibitem{Kazama:2014sxa}
Y.~Kazama, S.~Komatsu and T.~Nishimura, \emph{{Novel construction and the monodromy relation for three-point functions at weak coupling}}, \href{https://doi.org/10.1007/JHEP01(2015)095}{\emph{JHEP} {\bfseries 01} (2015) 095} [\href{https://arxiv.org/abs/1410.8533}{{\ttfamily 1410.8533}}].

\bibitem{Jiang:2015lda}
Y.~Jiang, S.~Komatsu, I.~Kostov and D.~Serban, \emph{{The hexagon in the mirror: the three-point function in the SoV representation}}, \href{https://doi.org/10.1088/1751-8113/49/17/174007}{\emph{J. Phys. A} {\bfseries 49} (2016) 174007} [\href{https://arxiv.org/abs/1506.09088}{{\ttfamily 1506.09088}}].

\bibitem{Kazama:2015iua}
Y.~Kazama, S.~Komatsu and T.~Nishimura, \emph{{On the singlet projector and the monodromy relation for psu(2, 2|4) spin chains and reduction to subsectors}}, \href{https://doi.org/10.1007/JHEP09(2015)183}{\emph{JHEP} {\bfseries 09} (2015) 183} [\href{https://arxiv.org/abs/1506.03203}{{\ttfamily 1506.03203}}].

\bibitem{Kazama:2016cfl}
Y.~Kazama, S.~Komatsu and T.~Nishimura, \emph{{Classical integrability for three-point functions: cognate structure at weak and strong couplings}}, \href{https://doi.org/10.1007/JHEP10(2016)042}{\emph{JHEP} {\bfseries 10} (2016) 042} [\href{https://arxiv.org/abs/1603.03164}{{\ttfamily 1603.03164}}].

\bibitem{Cavaglia:2018lxi}
A.~Cavagli\`a, N.~Gromov and F.~Levkovich-Maslyuk, \emph{{Quantum spectral curve and structure constants in $ \mathcal{N}=4 $ SYM: cusps in the ladder limit}}, \href{https://doi.org/10.1007/JHEP10(2018)060}{\emph{JHEP} {\bfseries 10} (2018) 060} [\href{https://arxiv.org/abs/1802.04237}{{\ttfamily 1802.04237}}].

\bibitem{Giombi:2018qox}
S.~Giombi and S.~Komatsu, \emph{{Exact Correlators on the Wilson Loop in $\mathcal{N}=4$ SYM: Localization, Defect CFT, and Integrability}}, \href{https://doi.org/10.1007/JHEP05(2018)109}{\emph{JHEP} {\bfseries 05} (2018) 109} [\href{https://arxiv.org/abs/1802.05201}{{\ttfamily 1802.05201}}].

\bibitem{Cavaglia:2021mft}
A.~Cavagli\`a, N.~Gromov and F.~Levkovich-Maslyuk, \emph{{Separation of variables in AdS/CFT: functional approach for the fishnet CFT}}, \href{https://doi.org/10.1007/JHEP06(2021)131}{\emph{JHEP} {\bfseries 06} (2021) 131} [\href{https://arxiv.org/abs/2103.15800}{{\ttfamily 2103.15800}}].

\bibitem{Bercini:2022jxo}
C.~Bercini, A.~Homrich and P.~Vieira, \emph{{Structure constants in N=4 supersymmetric Yang-Mills theory and separation of variables}}, \href{https://doi.org/10.1103/PhysRevD.110.L121901}{\emph{Phys. Rev. D} {\bfseries 110} (2024) L121901} [\href{https://arxiv.org/abs/2210.04923}{{\ttfamily 2210.04923}}].

\bibitem{Derkachov:2018rot}
S.~Derkachov, V.~Kazakov and E.~Olivucci, \emph{{Basso-Dixon Correlators in Two-Dimensional Fishnet CFT}}, \href{https://doi.org/10.1007/JHEP04(2019)032}{\emph{JHEP} {\bfseries 04} (2019) 032} [\href{https://arxiv.org/abs/1811.10623}{{\ttfamily 1811.10623}}].

\bibitem{Derkachov:2019tzo}
S.~Derkachov and E.~Olivucci, \emph{{Exactly solvable magnet of conformal spins in four dimensions}}, \href{https://doi.org/10.1103/PhysRevLett.125.031603}{\emph{Phys. Rev. Lett.} {\bfseries 125} (2020) 031603} [\href{https://arxiv.org/abs/1912.07588}{{\ttfamily 1912.07588}}].

\bibitem{Derkachov:2021rrf}
S.~Derkachov and E.~Olivucci, \emph{{Conformal quantum mechanics \& the integrable spinning Fishnet}}, \href{https://doi.org/10.1007/JHEP11(2021)060}{\emph{JHEP} {\bfseries 11} (2021) 060} [\href{https://arxiv.org/abs/2103.01940}{{\ttfamily 2103.01940}}].

\bibitem{Derkachov:2021ufp}
S.~Derkachov, G.~Ferrando and E.~Olivucci, \emph{{Mirror channel eigenvectors of the d-dimensional fishnets}}, \href{https://doi.org/10.1007/JHEP12(2021)174}{\emph{JHEP} {\bfseries 12} (2021) 174} [\href{https://arxiv.org/abs/2108.12620}{{\ttfamily 2108.12620}}].

\bibitem{Frenkel:1995zp}
E.~Frenkel, \emph{{Affine algebras, Langlands duality and Bethe ansatz}},  in \emph{{11th International Conference on Mathematical Physics (ICMP-11) (Satellite colloquia: New Problems in the General Theory of Fields and Particles, Paris, France, 25-28 Jul 1994)}}, 6, 1995 [\href{https://arxiv.org/abs/q-alg/9506003}{{\ttfamily q-alg/9506003}}].

\bibitem{Nekrasov:2009rc}
N.A.~Nekrasov and S.L.~Shatashvili, \emph{{Quantization of Integrable Systems and Four Dimensional Gauge Theories}},  in \emph{{16th International Congress on Mathematical Physics}}, pp.~265--289, 2010, \href{https://doi.org/10.1142/9789814304634_0015}{DOI} [\href{https://arxiv.org/abs/0908.4052}{{\ttfamily 0908.4052}}].

\bibitem{Frenkel:2015rda}
E.~Frenkel, S.~Gukov and J.~Teschner, \emph{{Surface Operators and Separation of Variables}}, \href{https://doi.org/10.1007/JHEP01(2016)179}{\emph{JHEP} {\bfseries 01} (2016) 179} [\href{https://arxiv.org/abs/1506.07508}{{\ttfamily 1506.07508}}].

\bibitem{Teschner:2017djr}
J.~Teschner, \emph{{Quantisation conditions of the quantum Hitchin system and the real geometric Langlands correspondence}},  \href{https://arxiv.org/abs/1707.07873}{{\ttfamily 1707.07873}}.

\bibitem{Jeong:2024onv}
S.~Jeong and N.~Lee, \emph{{Bispectral duality and separation of variables from surface defect transition}}, \href{https://doi.org/10.1007/JHEP12(2024)142}{\emph{JHEP} {\bfseries 12} (2024) 142} [\href{https://arxiv.org/abs/2402.13889}{{\ttfamily 2402.13889}}].

\bibitem{Faddeev:1996iy}
L.D.~Faddeev, \emph{{How algebraic Bethe ansatz works for integrable model}},  in \emph{{Les Houches School of Physics: Astrophysical Sources of Gravitational Radiation}}, pp.~pp. 149--219, 5, 1996 [\href{https://arxiv.org/abs/hep-th/9605187}{{\ttfamily hep-th/9605187}}].

\bibitem{Slavnov:2019hdn}
N.A.~Slavnov, \emph{{Introduction to the nested algebraic Bethe ansatz}}, \href{https://doi.org/10.21468/SciPostPhysLectNotes.19}{\emph{SciPost Phys. Lect. Notes} {\bfseries 19} (2020) 1} [\href{https://arxiv.org/abs/1911.12811}{{\ttfamily 1911.12811}}].

\bibitem{Slavnov:2018kfx}
N.A.~Slavnov, \emph{{Algebraic Bethe ansatz}} (4, 2018), \href{https://doi.org/10.1142/12776}{10.1142/12776}, [\href{https://arxiv.org/abs/1804.07350}{{\ttfamily 1804.07350}}].

\bibitem{Levkovich-Maslyuk:2016kfv}
F.~Levkovich-Maslyuk, \emph{{The Bethe ansatz}}, \href{https://doi.org/10.1088/1751-8113/49/32/323004}{\emph{J. Phys. A} {\bfseries 49} (2016) 323004} [\href{https://arxiv.org/abs/1606.02950}{{\ttfamily 1606.02950}}].

\bibitem{Gromov:2017blm}
N.~Gromov, \emph{{Introduction to the Spectrum of $N=4$ SYM and the Quantum Spectral Curve}},  \href{https://arxiv.org/abs/1708.03648}{{\ttfamily 1708.03648}}.

\bibitem{Levkovich-Maslyuk:2019awk}
F.~Levkovich-Maslyuk, \emph{{A review of the AdS/CFT Quantum Spectral Curve}}, \href{https://doi.org/10.1088/1751-8121/ab7137}{\emph{J. Phys. A} {\bfseries 53} (2020) 283004} [\href{https://arxiv.org/abs/1911.13065}{{\ttfamily 1911.13065}}].

\bibitem{Kazakov:2018ugh}
V.~Kazakov, \emph{{Quantum Spectral Curve of $\gamma$-twisted ${\cal N}=4$ SYM theory and fishnet CFT}},  \href{https://arxiv.org/abs/1802.02160}{{\ttfamily 1802.02160}}.

\bibitem{Chernyak:2020lgw}
D.~Chernyak, S.~Leurent and D.~Volin, \emph{{Completeness of Wronskian Bethe Equations for Rational ${\mathfrak {\mathfrak {gl}}_{{{\mathsf {m}}}|{{\mathsf {n}}}}}$ Spin Chains}}, \href{https://doi.org/10.1007/s00220-021-04275-9}{\emph{Commun. Math. Phys.} {\bfseries 391} (2022) 969} [\href{https://arxiv.org/abs/2004.02865}{{\ttfamily 2004.02865}}].

\bibitem{Kazama:2013rya}
Y.~Kazama, S.~Komatsu and T.~Nishimura, \emph{{A new integral representation for the scalar products of Bethe states for the XXX spin chain}}, \href{https://doi.org/10.1007/JHEP09(2013)013}{\emph{JHEP} {\bfseries 09} (2013) 013} [\href{https://arxiv.org/abs/1304.5011}{{\ttfamily 1304.5011}}].

\bibitem{sklyanin1989new}
E.~Sklyanin, \emph{New approach to the quantum nonlinear schrodinger equation}, {\emph{Journal of Physics A: Mathematical and General} {\bfseries 22} (1989) 3551}.

\bibitem{Belliard:2018pie}
S.~Belliard and N.A.~Slavnov, \emph{{A note on $\mathfrak{gl}_2$-invariant Bethe vectors}}, \href{https://doi.org/10.1007/JHEP04(2018)031}{\emph{JHEP} {\bfseries 04} (2018) 031} [\href{https://arxiv.org/abs/1802.07576}{{\ttfamily 1802.07576}}].

\bibitem{Belliard:2018pvg}
S.~Belliard, N.A.~Slavnov and B.~Vallet, \emph{{Modified Algebraic Bethe Ansatz: Twisted XXX Case}}, \href{https://doi.org/10.3842/SIGMA.2018.054}{\emph{SIGMA} {\bfseries 14} (2018) 054} [\href{https://arxiv.org/abs/1804.00597}{{\ttfamily 1804.00597}}].

\bibitem{molev2007yangians}
A.~Molev, \emph{Yangians and classical Lie algebras}, no.~143, American Mathematical Soc. (2007).

\bibitem{borot2016asymptotic}
G.~Borot, A.~Guionnet and K.K.~Kozlowski, \emph{Asymptotic expansion of a partition function related to the sinh-model}, Springer (2016).

\bibitem{borot2015large}
G.~Borot, A.~Guionnet and K.K.~Kozlowski, \emph{Large-n asymptotic expansion for mean field models with coulomb gas interaction}, {\emph{International Mathematics Research Notices} {\bfseries 2015} (2015) 10451}.

\bibitem{keating2004random}
J.P.~Keating and F.~Mezzadri, \emph{Random matrix theory and entanglement in quantum spin chains}, {\emph{Communications in mathematical physics} {\bfseries 252} (2004) 543}.

\bibitem{Castro-Alvaredo:2010scy}
O.A.~Castro-Alvaredo and B.~Doyon, \emph{{Permutation operators, entanglement entropy, and the XXZ spin chain in the limit \textbackslash{}Delta -\ensuremath{>} -1}}, \href{https://doi.org/10.1088/1742-5468/2011/02/P02001}{\emph{J. Stat. Mech.} {\bfseries 1102} (2011) P02001} [\href{https://arxiv.org/abs/1011.4706}{{\ttfamily 1011.4706}}].

\bibitem{refael2004entanglement}
G.~Refael and J.E.~Moore, \emph{Entanglement entropy of random quantum critical points in one dimension}, {\emph{Physical review letters} {\bfseries 93} (2004) 260602}.

\bibitem{Castro-Alvaredo:2016cdj}
O.A.~Castro-Alvaredo, B.~Doyon and T.~Yoshimura, \emph{{Emergent hydrodynamics in integrable quantum systems out of equilibrium}}, \href{https://doi.org/10.1103/PhysRevX.6.041065}{\emph{Phys. Rev. X} {\bfseries 6} (2016) 041065} [\href{https://arxiv.org/abs/1605.07331}{{\ttfamily 1605.07331}}].

\bibitem{DeNardis:2018omc}
J.~De~Nardis, D.~Bernard and B.~Doyon, \emph{{Hydrodynamic Diffusion in Integrable Systems}}, \href{https://doi.org/10.1103/PhysRevLett.121.160603}{\emph{Phys. Rev. Lett.} {\bfseries 121} (2018) 160603} [\href{https://arxiv.org/abs/1807.02414}{{\ttfamily 1807.02414}}].

\bibitem{McGovern:2019sdd}
J.~McGovern, \emph{{Scalar insertions in cusped Wilson loops in the ladders limit of planar $ \mathcal{N} $ = 4 SYM}}, \href{https://doi.org/10.1007/JHEP05(2020)062}{\emph{JHEP} {\bfseries 05} (2020) 062} [\href{https://arxiv.org/abs/1912.00499}{{\ttfamily 1912.00499}}].

\bibitem{Caetano:2020dyp}
J.a.~Caetano and S.~Komatsu, \emph{{Functional equations and separation of variables for exact $g$-function}}, \href{https://doi.org/10.1007/JHEP09(2020)180}{\emph{JHEP} {\bfseries 09} (2020) 180} [\href{https://arxiv.org/abs/2004.05071}{{\ttfamily 2004.05071}}].

\bibitem{Gurdogan:2015csr}
O.~G\"urdo\u{g}an and V.~Kazakov, \emph{{New Integrable 4D Quantum Field Theories from Strongly Deformed Planar $\mathcal N = $ 4 Supersymmetric Yang-Mills Theory}}, \href{https://doi.org/10.1103/PhysRevLett.117.201602}{\emph{Phys. Rev. Lett.} {\bfseries 117} (2016) 201602} [\href{https://arxiv.org/abs/1512.06704}{{\ttfamily 1512.06704}}].

\bibitem{Derkachov:2018ewi}
S.E.~Derkachov and P.A.~Valinevich, \emph{{Separation of variables for the quantum $SL(3,\mathbb C)$ spin magnet: eigenfunctions of Sklyanin $B$-operator}}, \href{https://doi.org/10.1007/s10958-019-04505-5}{\emph{Zap. Nauchn. Semin.} {\bfseries 473} (2018) 110} [\href{https://arxiv.org/abs/1807.00302}{{\ttfamily 1807.00302}}].

\bibitem{Chicherin:2017cns}
D.~Chicherin, V.~Kazakov, F.~Loebbert, D.~M\"uller and D.-l.~Zhong, \emph{{Yangian Symmetry for Bi-Scalar Loop Amplitudes}}, \href{https://doi.org/10.1007/JHEP05(2018)003}{\emph{JHEP} {\bfseries 05} (2018) 003} [\href{https://arxiv.org/abs/1704.01967}{{\ttfamily 1704.01967}}].

\bibitem{Chicherin:2017frs}
D.~Chicherin, V.~Kazakov, F.~Loebbert, D.~M\"uller and D.-l.~Zhong, \emph{{Yangian Symmetry for Fishnet Feynman Graphs}}, \href{https://doi.org/10.1103/PhysRevD.96.121901}{\emph{Phys. Rev. D} {\bfseries 96} (2017) 121901} [\href{https://arxiv.org/abs/1708.00007}{{\ttfamily 1708.00007}}].

\bibitem{Loebbert:2022nfu}
F.~Loebbert, \emph{{Integrability for Feynman integrals}}, \href{https://doi.org/10.21468/SciPostPhysProc.14.008}{\emph{SciPost Phys. Proc.} {\bfseries 14} (2023) 008} [\href{https://arxiv.org/abs/2212.09636}{{\ttfamily 2212.09636}}].

\bibitem{Kazakov:2023nyu}
V.~Kazakov, F.~Levkovich-Maslyuk and V.~Mishnyakov, \emph{{Integrable Feynman Graphs and Yangian Symmetry on the Loom}},  \href{https://arxiv.org/abs/2304.04654}{{\ttfamily 2304.04654}}.

\bibitem{Levkovich-Maslyuk:2024zdy}
F.~Levkovich-Maslyuk and V.~Mishnyakov, \emph{{Yangian symmetry, GKZ equations and integrable Feynman graphs in conformal variables}},  \href{https://arxiv.org/abs/2412.19296}{{\ttfamily 2412.19296}}.

\bibitem{Buric:2020dyz}
I.~Buric, S.~Lacroix, J.A.~Mann, L.~Quintavalle and V.~Schomerus, \emph{{From Gaudin Integrable Models to $d$-dimensional Multipoint Conformal Blocks}}, \href{https://doi.org/10.1103/PhysRevLett.126.021602}{\emph{Phys. Rev. Lett.} {\bfseries 126} (2021) 021602} [\href{https://arxiv.org/abs/2009.11882}{{\ttfamily 2009.11882}}].

\bibitem{Buric:2021ywo}
I.~Buric, S.~Lacroix, J.A.~Mann, L.~Quintavalle and V.~Schomerus, \emph{{Gaudin models and multipoint conformal blocks: general theory}}, \href{https://doi.org/10.1007/JHEP10(2021)139}{\emph{JHEP} {\bfseries 10} (2021) 139} [\href{https://arxiv.org/abs/2105.00021}{{\ttfamily 2105.00021}}].

\bibitem{Arnaudon:2004sd}
D.~Arnaudon, J.~Avan, N.~Crampe, A.~Doikou, L.~Frappat and E.~Ragoucy, \emph{{General boundary conditions for the sl(N) and sl(M|N) open spin chains}}, \href{https://doi.org/10.1088/1742-5468/2004/08/P08005}{\emph{J. Stat. Mech.} {\bfseries 0408} (2004) P08005} [\href{https://arxiv.org/abs/math-ph/0406021}{{\ttfamily math-ph/0406021}}].

\bibitem{Belliard:2019bfz}
S.~Belliard and N.A.~Slavnov, \emph{{Why scalar products in the algebraic Bethe ansatz have determinant representation}}, \href{https://doi.org/10.1007/JHEP10(2019)103}{\emph{JHEP} {\bfseries 10} (2019) 103} [\href{https://arxiv.org/abs/1908.00032}{{\ttfamily 1908.00032}}].

\end{thebibliography}\endgroup

\end{document}